\documentclass[fleqn,usenatbib,useAMS]{mnras}

\usepackage{graphicx}   
\usepackage{amsmath}    
\usepackage{amssymb}    
\usepackage{multicol}        
\usepackage{bm}         
\usepackage{pdflscape}  

\usepackage[T1]{fontenc}
\usepackage{ae,aecompl}

\usepackage{newtxtext,newtxmath}

\usepackage[usenames,dvipsnames]{xcolor}

\usepackage{soul}

\def\d{$^\circ$}
\def\m{$^\prime$}
\def\s{$^{\prime\prime}$}

\title[White dwarf - red dwarf orbital periods]{Evidence for bimodal orbital separations of white dwarf-red dwarf binary stars }

\author[R.P. Ashley et al.]{R.P. Ashley$^{1}$\thanks{E-mail: r.p.ashley@warwick.ac.uk},
J. Farihi$^{2}$,
T.R. Marsh$^{1}$
D.J. Wilson$^{1}$ and 
B.T. G{\"a}nsicke$^{1}$,
\\
$^{1}$Department of Physics, University of Warwick, Gibbet Hill Road, Coventry, CV4 7AL, UK\\
$^{2}$Department of Physics and Astronomy, University College London, London WC1E 6BT\\
}

\date{Accepted XXX. Received YYY; in original form ZZZ}

\pubyear{2017}

\begin{document}
\label{firstpage}
\pagerange{\pageref{firstpage}--\pageref{lastpage}}
\maketitle

\begin{abstract}
We present the results of a radial velocity survey of 20 white dwarf plus M dwarf binaries selected as a follow up to a \textit{Hubble Space Telescope} study that aimed to spatially resolve suspected binaries. Our candidates are taken from the list of targets that were spatially unresolved with \textit{Hubble}. We have determined the orbital periods for 16 of these compact binary candidates. The period distribution ranges from 0.14 to 9.16\,d and peaks near 0.6\,d. The original sample therefore contains two sets of binaries, wide orbits ($\approx100-1000$\,au) and close orbits ($\lesssim1-10$\,au), with no systems found in the $\approx10-100$\,au range. This observational evidence confirms the bimodal distribution predicted by population models and is also similar to results obtained in previous studies. We find no binary periods in the months to years range, supporting the post common envelope evolution scenario.  One of our targets, WD\,1504+546, was discovered to be an eclipsing binary with a period of 0.93\,d.  
\end{abstract}



\section{Introduction}
It is estimated that the field star population in the solar neighbourhood consists of 50 per\,cent binary systems \citep{binaryfraction}. During the evolutionary lifetime of these systems many can evolve into compact binaries. Compact binaries are some of the most interesting objects in the Galaxy, and include the precursors to type Ia supernovae that are the cosmic yardsticks used to measure inter-galactic distances and, consequently, the scale of the universe. Compact binary evolution is also the path that leads to the formation of a pair of neutron stars that, when merging, produce the gravitational wave signals that are now being detected by LIGO and VIRGO. Other examples are X-ray binaries, double-degenerate binaries \citep{Nelemans2005}, planetary nebulae \citep{DeMarco2009} and cataclysmic variables \citep{Tappert2009}. 
%


A fundamental phase in the evolution of these systems is the \emph{common envelope}. After their initial formation, both components in a binary system live out their main sequence lives without interaction, but when the more massive component, the primary star, leaves the main sequence, it will evolve up the red giant branch. If the separation of the components in the binary is sufficiently small, the primary star's increase in size will be such that it fills the Roche lobe. At this point unstable mass transfer is triggered and the primary's atmosphere will extend sufficiently far to engulf both stars and thereby form a common envelope. During the common envelope phase, drag forces will drain angular momentum from both bodies and bring them closer. This angular momentum is transferred to the gas envelope and helps to expel it from the system. The exact processes that occur during the common envelope phase are poorly understood, and difficult to model computationally, but it is thought that the inspiral occurs quickly \cite[$<100$ years,][]{CEreview}. Eventually, the primary will evolve away from its giant phase, leaving behind a white dwarf. The end result of this process is a pair consisting of a white dwarf and a main sequence star (WD+MS) with a relatively small separation, known as a post common-envelope binary (PCEB). However, not all binary systems will follow this path. If the initial separation of the two components is larger than about $2-3$\,au \citep{WillemsKolb}, then the red giant star does not fill its Roche lobe and no common envelope is formed. In fact, when the outer atmosphere of the primary is dissipated, this mass loss will have the effect of widening the orbit. These two distinct evolutionary paths lead to two separate populations of WD+MS binaries, those that have experienced a common envelope and therefore have periods of hours to days, and those that have not experienced a common envelope and have periods of several years or more. 

A commonly used method to investigate the outcome of the common envelope phase is Binary Population Synthesis (BPS). This is a theoretical process that involves simulating the evolution of a binary by modeling the development of both components using stellar evolution models and including additional simulations of mass transfer and orbital dynamics \citep{WillemsKolb}. BPS studies have been performed for Type~Ia supernova progenitors \citep{2004MNRAS.350.1301H}, short $\gamma$-ray bursts \citep{2006ApJ...648.1110B}, and, most useful for this study, WD+MS systems \citep{WillemsKolb, Davis2010}. These studies predict that WD+MS systems will have orbital periods that follow a bimodal distribution, with the short period PCEBs peaking at about $1-2$\,d ($\log P_{\rm orb}\,(\rm d)\approx 0.2$), and continuously detached binaries with periods of years or more. 



The fraction of the orbital energy that goes into discarding the common envelope is usually given as $\alpha_{\mathrm{CE}}$ where $E_{\mathrm{bind}} = \alpha_{\mathrm{CE}} \Delta E_{\mathrm{orb}}$. $E_{\mathrm{bind}}$ is the gravitational binding energy of the envelope and $\Delta E_{\mathrm{orb}}$ is the change in orbital energy of the two stars \citep{Davis2010, WillemsKolb, Zorotovic2014}. Recent observational and theoretical studies indicate a value for the efficiency of transfer of the orbital energy $\alpha_{\mathrm{CE}}=0.2-0.3$ \citep{Zorotovic2010}. 

An additional energy source that might be a factor in dispersing the common envelope is the recombination energy released when the ionised hydrogen gas in the common envelope is neutralised. \citet{Zorotovic2014} show that increasing the contribution that recombination energy plays in the dispersal of the envelope would lead to longer period PCEBs overall with a tail in the period distribution going out to about 1000 days, since this energy would contribute to the dispersal of the common envelope and less energy would be needed from the angular momentum loss.  At the moment no PCEB systems are known to have periods of this length. The longest known PCEB period is that of IK Peg at 21.72 days \citep{test}. 

In order to test the underlying period distribution of WD+MS binaries we need an observationally derived period distribution of an unbiased sample. In recent years, large scale surveys such as SDSS \citep{Sloan2015} have enabled the study of the orbital period distribution of a sample of WD+MS binaries and measurement of the fraction of those that have undergone a common envelope phase \citep{2012MNRAS.423..320R, Nebot2011, Schreiber2008, Schreiber2010}. These studies have shown broad agreement with predictions of the fraction of PCEBs versus widely separated binaries and the shape of the period distribution for the PCEBs themselves. \citet{Nebot2011} concluded that about a third of the WD+MS binaries identified by SDSS are short periods PCEBs, the rest are candidate wide binaries. These are systems that, at the SDSS imaging resolution, $\approx 1-2$\s, are still unresolved, but since they are at distances of a few 100\,pc, they could have separations of 100\,au or more.

In this study, we used spectroscopy to search for radial velocity variations in 20 WD+MS binaries to ascertain if they are PCEB systems.  We started with a sample created by \citet{Farihi2006} that were selected by searching for white dwarfs with near-infrared excess. Our targets were chosen from those that were shown to be unresolved in images taken by the \textit{Hubble Space Telescope} (\textit{HST}). In the context of post-common envelope evolution, theory predicts these systems should all be close binaries. 

\section{Observations} \label{observations}

\subsection{Target selection} \label{targetselection}
\citet{FarihiFull} obtained \textit{HST} high-resolution imaging of WD+MS binaries to test the bimodal period distribution with an unbiased sample. The targets selected were chosen from the \citet{McCookSion1999} catalogue of white dwarfs which were found to exhibit a near-infrared excess in 2MASS photometry by \citet{Wachter2003}. A complete list of the original targets can be found in Table 1 of \citet{Farihi2006}. 

The Advanced Camera for Surveys (ACS) High-Resolution Camera (HRC) was used to image these targets with the F814W filter to try to resolve each suspected binary. The results of this exercise can be seen in Table 7 of \citet{FarihiFull}, where it was found that 72 of the 90 candidates were highly probable binaries and, of these, 43 systems were spatially resolved with the ACS HRC. Combining the angular separation with distance estimates to each target allowed \citet{FarihiFull} to compute the minimum value for the semi-major axis (at zero inclination) for each binary pair. 

The remaining 29 systems were unresolved at the resolution of the \textit{HST} imaging. Taking the spatially resolved limit of one pixel on the ACS HRC camera (0.025 arcsec) as the resolving power of the imaging study, it was possible to place upper limits on the apparent separation on these unresolved targets. At the distances to these targets, the limits are about 10\,au, corresponding to orbital periods of tens of years. In this study, we have attempted to obtain radial velocity measurements of these systems.  

We obtained spectroscopy of 20 of the 29 WD+MS binaries that were accessible from the Isaac Newton Telescope at the Roque de los Muchachos Observatory, Canary Islands, Spain. To constrain and potentially determine the orbital periods, the observations were taken in the \textit{I}\,-band (6500$-$9000\,\AA) region to measure features in the atmosphere of the secondary, and monitor them for radial velocity variations. The \ion{Na}{i} doublet at 8183.3 and 8194.8\,{\AA} and the H\,$\alpha$ line at 6562.8\,{\AA} were used. 

As a followup to our spectroscopy, we examined the archives of the Catalina Real-time Transient Survey (CRTS) \citep{CatalinaCatalog} and the Palomar Transient Factory (PTF) \citep{PTFCatalog} to look for photometric observations for all of our targets. We discovered that of our 20 systems, 16 had CRTS photometry and 10 had PTF photometry. In the cases of WD\,0303$-$007, WD\,1458$+$171, and WD\,1504+546, these data provided additional insight and this is discussed in section \ref{individualnotes}. 

\subsection{Reduction}
The spectra were obtained using the Intermediate Dispersion Spectrograph (IDS) with the R831R grating. This is a medium dispersion grating with a dispersion of 0.75\,{\AA} per pixel and a resolving power of $R=4667$ at 7000\,\AA. The central wavelengths were chosen to ensure that features of interest were placed near the centre of the detector, 8200\,{\AA} for the \ion{Na}{i} doublet and 6560\,{\AA} for H\,$\alpha$. A summary of our observations is shown in Table \ref{tab:targets}.

Each observing night included at least two standard stars in order to enable flux calibration and the removal of telluric features. All of the spectra were optimally extracted and reduced using the {\sc pamela} and {\sc molly} reduction software, \citep{MarshPamela}. Our comparison stars were chosen from early-type main-sequence stars and white dwarfs since their spectra have few features in the $7500-8700$\,{\AA} region. A list of these standards is given in Table \ref{tab:standards}. We fitted splines to the continua of our comparison stars, avoiding regions of atmospheric absorption from 7560 to 7690\,{\AA} and 8100 to 8400\,\AA. The ratio of the fit to the original spectrum was then used to correct for the absorption in the target spectra \cite[see  e.g.][]{Marsh1990}. The calibration star with the nearest air-mass was used and the correction was scaled by the ratio of the airmasses to the power 0.55 to account for the saturated absorption bands. The removal of these significant telluric features was not always adequate and this meant that some spectra could not be used for radial velocity fits as the \ion{Na}{i} features were not visible after this correction.    

For wavelength calibration, CuNe and CuAr arcs were observed several times over the course of the observations. This was then further improved by examining sky emission features in each of the science and calibration spectra and calculating the final wavelength shift required to assign these features to their known wavelengths. The sky emission lines used were the oxygen emission features at 7913.7 and 8344.6\,\AA. 

\begin{table}
  \caption{The targets observed during the campaign. The coordinates are in epoch J2000. Spectra were deemed unusable if we were unable to fit a profile to the \ion{Na}{i} or H\,$\alpha$ feature. For all targets the \ion{Na}{i} doublet at 8190\,{\AA} was observed, with the exception of WD\,2257+162, where we observed the H\,$\alpha$ emission from the secondary. }
  \begin{tabular}{ c  c  c  c  c }
    \hline
    WD & RA       & Dec            & \# Spectra & \# Usable  \\
       & (h m s)  & (\d~ \m~ \s)   & taken   & spectra \\
  \hline
    0023$+$388 & 00 26 33.2 & +39 09 03 & 21 & 15 \\
    0303$-$007 & 03 06 07.1 & $-$00 31 14 & 31 & 31 \\
    0354$+$463 & 03 58 17.1 & +46 28 41 & 57 & 46 \\
    0430$+$136 & 04 33 10.3 & +13 45 17 & 30 & 29 \\
    0752$-$146 & 07 55 08.9 & $-$14 45 53 & 30 & 20 \\
    0812$+$478 & 08 15 48.9 & +47 40 39 & 16 & 16 \\
    0908$+$226 & 09 11 43.1 & +22 27 49 & 26 & 23 \\
    1001$+$203 & 10 04 04.2 & +20 09 23 & 22 & 22 \\
    1037$+$512 & 10 40 16.8 & +50 56 47 & 22 & 22 \\
    1051$+$516 & 10 54 21.9 & +51 22 54 & 22 & 22 \\
    1133$+$358 & 11 35 42.7 & +35 34 24 & 19 & 19 \\
    1333$+$487 & 13 36 01.7 & +48 28 45 & 17 & 8 \\
    1339$+$606 & 13 41 00.0 & +60 26 10 & 20 & 12 \\
    1433$+$538 & 14 34 43.1 & +53 35 25 & 26 & 22 \\
    1436$-$216 & 14 39 12.8 & $-$21 50 12 & 32 & 32 \\
    1458$+$171 & 15 00 19.4 & +16 59 16 & 18 & 10 \\
    1504$+$546 & 15 06 05.3 & +54 28 19 & 39 & 39 \\
    1517$+$502 & 15 19 05.9 & +50 07 03 & 22 & 0 \\
    2257$+$162 & 22 59 46.9 & +16 29 17 & 16 & 14 \\
    2317$+$268 & 23 20 04.1 & +25 52 21 & 12 & 11 \\    
    \hline
  \end{tabular}
  \label{tab:targets}
\end{table}

\section{Analysis} \label{analysis}
Radial velocities for all spectra were calculated by least-squares fits of a double-Gaussian function to the data near the \ion{Na}{i} doublet at 8190\,\AA. The separation of the doublet was kept fixed at the laboratory value of 11.5\,\AA; the depth of the lines were free to vary but kept equal to each other, and the width of the lines was fixed to a FWHM of 2.4\,\AA. One of the targets, WD\,1001+203, showed consistently broad lines on all of the spectra, so for this object, a larger width of 4.7\,{\AA} was used. The uncertainties of the fit parameters were taken from the co-variance matrix produced in the Levenberg-Marquardt procedure. The fitted wavelengths were then converted to radial velocities by calculating the shift to the rest wavelength for the blue-ward line of the doublet, 8183.3\,\AA.  

 
For the target WD\,2257+162, it was noted that previous observations had been unable to detect the \ion{Na}{i} doublet, \citep{Liebert2005, Tremblay2007}, and therefore spectra centred on the H\,$\alpha$ line were taken. Two overlapping Gaussian functions, one representing the emission profile and a second for the absorption profile were fitted. Radial velocities were derived by comparing the wavelength to the rest wavelength of H\,$\alpha$. While the fit to the emission gave errors of about 10 km\,s$^{-1}$, similar to measurements of the \ion{Na}{i} doublet for the other targets, the absorption feature was too broad to allow the determination of the radial velocity of the white dwarf. The measured radial velocities for all of the targets are listed in Table \ref{tab:spectroscopy} of the Appendix.

\subsection{Binary periods}  \label{periods}
In order to identify which of the targets exhibited radial velocity variability we used a test described in \citet{Maxted2008}, where we calculate the probability, $p$ of obtaining the computed $\chi^2$ value compared to that computed from a sample of measurements distributed normally around the mean radial velocity. We set the criterion for radial velocity detection to $\log(p) < -4$. The three targets that did not meet this criterion were WD\,0430+221, WD\,0812+478, and WD\,1001+203. We were unable to detect any \ion{Na}{i} absorption lines in our spectra of WD\,1517+302 and therefore could not deduce any radial velocity variability for this system.  

Periods were derived by fitting a sine wave plus a constant to the data, and choosing the frequency that corresponded to the lowest $\chi^2$. Our algorithm used the floating mean periodogram approach \citep{1999ApJ...526..890C}. The key point in this method is that the constant systemic velocity is fitted at the same time as semi-amplitude and phase. This corrects a failing of the well-known Lomb-Scargle \citep{Lomb1976, Scargle1982} periodogram which starts by subtracting the mean of the data and then fits a plain sinusoid and this is incorrect for small numbers of points.

The data sampling that occurred as a result of our observing strategy, i.e.  taking a spectrum once or twice per night during a week long run, means that we were susceptible to aliasing, especially at multiples of $1\,\mathrm{d}^{-1}$. We also had long gaps of several months between observing runs, which caused a fine-splitting of the 1\,d aliases.  Although it is certain that the radial velocities measured are due orbital motion and also that the periods are relatively short (on the order of days), in some cases there are several competing aliases. Table \ref{tab:periods} lists the best period and the most significant competing alias.  Fig. \ref{fig:rvplots} shows the fitted radial velocity plots and periodograms for all 19 of the objects that had fitted radial velocity data. This figure includes the three targets,  WD\,0430+136, WD\,0812+478, and WD\,1001+203 that failed the radial velocity variability test but are shown here for completeness. 

\begin{table}
  \caption{Fitted periods and aliases shown with the corresponding reduced $\chi^2$ value. For each target, the number of valid radial velocity measurements $N$ used in the analysis is also shown. We show the next best competing alias along with the associated increase in $\chi^2$.}
 \begin{tabular}[width=\textwidth]{ l  l  l  l  l  l  }
    \hline
    WD & $N$  &  $P_{\rm orb}$  & $\chi^2_{\mathrm{red}}$ & $2^{\rm nd}$ best alias & $\Delta \chi^2$ \\
       &    &  (d)  &		  		& (d)		 &                 \\
    \hline 
     0023$+$388        & 19 & 0.64159(1)  & 1.2    & 0.68339(1)  & 12 \\
     0303$-$007        & 31 & 0.54113(1)  & 1.0    & 1.19096(4)  & 135 \\ 
     0354$+$463        & 46 & 0.165203(1) & 1.0    & 0.246896(2) & 154 \\
     0430$+$136\textsuperscript{*}       & 30 & 0.135480(7) & 0.2    & 0.119120(6) & 0 \\
     0752$-$146        & 21 & 1.05241(3)  & 1.0    & 0.522410(5) & 15 \\
     0812$+$478\textsuperscript{*}       & 16 & 0.055373(1) & 0.2    & 0.7141(2)   & 0  \\
     0908$+$226        & 22 & 9.1614(8)   & 0.5    & 9.393(8)    & 2 \\
     1001$+$203\textsuperscript{*}       & 22 & 0.26579(2)  & 0.2    & 0.20913(2)  & 0 \\
     1037$+$512        & 26 & 0.141460(1) & 0.5    & 0.165146(2) & 1 \\
     1051$+$516        & 22 & 2.770208(3) & 0.5    & 0.73322(2)  & 16 \\
     1133$+$358        & 18 & 5.956(2)    & 0.0    & 6.055(2)    & 0 \\
     1333$+$487        & 11 & 1.75584(2)  & 0.0    & 2.27544(7)  & 1 \\
     1339$+$606        & 17 & 0.49367(1)  & 0.4    & 0.247534(4) & 0 \\
     1433$+$538        & 22 & 4.479(3)    & 0.4    & 4.357(2)    & 0 \\
     1436$-$216        & 32 & 2.01774(2)  & 0.4    & 0.66988(4)  & 5 \\
     1458$+$171        & 10 & 0.0793902(3)& 1.5    & 0.164702(1) & 5 \\
     1504$+$546        & 36 & 0.93072(2)  & 1.0    & 0.479524(5) & 128 \\ 
     2257$+$162        & 15 & 0.3223(1)   & 0.7    & 0.4736(5)   & 21 \\ 
     2317$+$268        & 10 & 0.7944(9)   & 1.0    & 3.85(5)     & 0  \\ 
    \hline
  \end{tabular}
    \textbf{Notes.} (*) No significant radial velocity variation detected.
  \label{tab:periods}
\end{table}

Once the best frequency had been identified, it was used as a starting point for a least-squares fit of a sine wave to the data of the following form, 
\begin{equation}
	v_\mathrm{r} = \gamma_2 + K_2 \sin \Big [\frac{2\pi(t-t_0)}{P_{\rm orb}} \Big], 
\end{equation} 
where the initial guess for $P_{\rm orb}$ was $1/f$ and $f$ is the frequency at the lowest $\chi^2$ on the periodogram and allowed to vary in the fit, $\gamma_2$ is the systemic velocity of the secondary star, $t_0$ is the zero point defined by the inferior conjunction of the secondary star, and $K_2$ is the radial velocity semi-amplitude of the secondary star.

Figure \ref{fig:rvplots} shows the reduced $\chi^2$ value for each of the fitted radial velocity curves. Several systems showed clear signs of extra noise, with minimum $\chi^2$ values much larger than the number of degrees of freedom. Particularly discrepant spectra were examined for signs of problems, and the conclusion drawn was that the most likely cause was residual and/or poorly corrected telluric absorption. Non-uniform filling of the spectrograph slit could have also been a contributing factor. In order to account for this extra source of noise, a fixed value was added, in quadrature, to all uncertainties to make the reduced $\chi^2$ equal to unity for each fit. The typical fixed value added the uncertainties was 7--10~km\,s$^{-1}$. This corresponds to about one third of a pixel in the spectrograph dispersion. Experience shows that this is a little larger than typical fluctuations caused by non-uniform slit illumination, hence we suspect that telluric correction is the more significant issue. Although this noise reduces the precision of our data, the radial velocity signal in most of our targets leaves no doubt as to the reality of the variations.

For all targets, the $\chi^2$ value for the second best alias was also calculated. These values are listed in Table \ref{tab:periods}. It can be shown that the probability of a given period being the correct one in a Bayesian sense scales as $e^{-\chi^2/2}$ \citep{Marsh1995, Morales2003}. This means that for large values of $\Delta\chi^2$ we can be sure of our period, but for values of $\Delta\chi^2<5$ there is a significant chance of the correct period being one of the other aliases. Examination of Table \ref{tab:periods} indicates that, while we can be sure that WD\,0908$+$226, WD\,1037$+$512, WD\,1133$+$358, WD\,1333$+$487, WD\,1339$+$606, WD\,1433$+$53, and WD\,2317$+$268 exhibit radial velocity variability, we have not determined the true period with a high level of confidence. 

\begin{table*}
  \caption{Parameters for the binaries determined from the radial velocity study. The period, $K_2$ and $\gamma_2$ velocities are from this paper, the secondary estimated spectral type is taken from \citet{FarihiFull}, and the $T_{\rm eff}$ has been retrieved from the Montreal White Dwarf Database \citep{MontrealDB}. }
  \begin{tabular}[width=\textwidth]{ c  l  r  r  l  r  }
    \hline
    WD 			& $P_{\rm orb}$  & $K_2$           & $\gamma_2$      & SpType & $T_\mathrm{eff}$ \\
       			& (d)     & (km\,s$^{-1}$)  & (km\,s$^{-1}$)  &        & (K)               \\
    \hline
    0023$+$388  & 0.64159(1)   & 153(4) & --24(3) & DA+dM5.5    & 10 980  \\
    0303$-$007  & 0.54113(1)   & 132(3) & 6(2)    & DA+dM4      & 20 310  \\
    0354$+$463  & 0.165203(1)  & 115(3) & --67(2) & DA+dM7      &  8 230  \\
    0430$+$136  &              &        &         & DA+dM5.5    & 34 210  \\
    0752$-$146  & 1.05241(3)   & 159(10)& 0(13)   & DA+dM6      & 19 440  \\
    0812$+$478  &              &        &         & DA+dM4      & 62 000  \\
    0908$+$226  & 9.1614(8)    & 48(4)  & --35(4) & DA+dM3      & 10 548  \\
    1001$+$203  &    	       &        &         & DA+dM2.5    & 21 010  \\
    1037$+$512  & 0.141460(1)  & 18(4)  & --46(3) & DA+dM4      & 19 780  \\
    1051$+$516  & 2.770208(3)  & 81(3)  & 10(2)   & DA+dM3      & 23 863  \\
    1133$+$358  & 5.956(2)     & 78(3)  & --17(3) & DC+dM4.5    &  6 500  \\
    1333$+$487  & 1.75584(2)   & 153(9) & --81(9) & DB+dM6.5    & 14 676  \\
    1339$+$606  & 0.49367(1)   & 73(9)  & --2(6)  & DA+dM3.5    & 44 770  \\
    1433$+$538  & 4.479(3)     & 81(4)  & --44(3) & DA+dM4.5    & 23 260  \\
    1436$-$216  & 2.01774(2)   & 44(10) & --4(2)  & DA+dM2.5    & 23 690  \\
    1458$+$171  & 0.164702(1)  & 180(8) & 23(6)   & DA+dM4.5    & 22 600  \\
    1504$+$546  & 0.93072(2)   & 129(4) & --3(3)  & DA+dM3      & 23 120  \\
    1517$+$502  &              &        &         & DA+dC       & 31 270  \\
    2257$+$162  & 0.3223(1)    & 74(7)  & 16(5)   & DA+dM4.5    & 25 450  \\
    2317$+$268  & 0.7941(2)    & 125(6) & 12(4)   & DA+dM3.5    & 31 890  \\
    \hline
  \end{tabular}
  \label{tab:parameters}
\end{table*}

\begin{figure}
\centering
\includegraphics[width=\columnwidth]{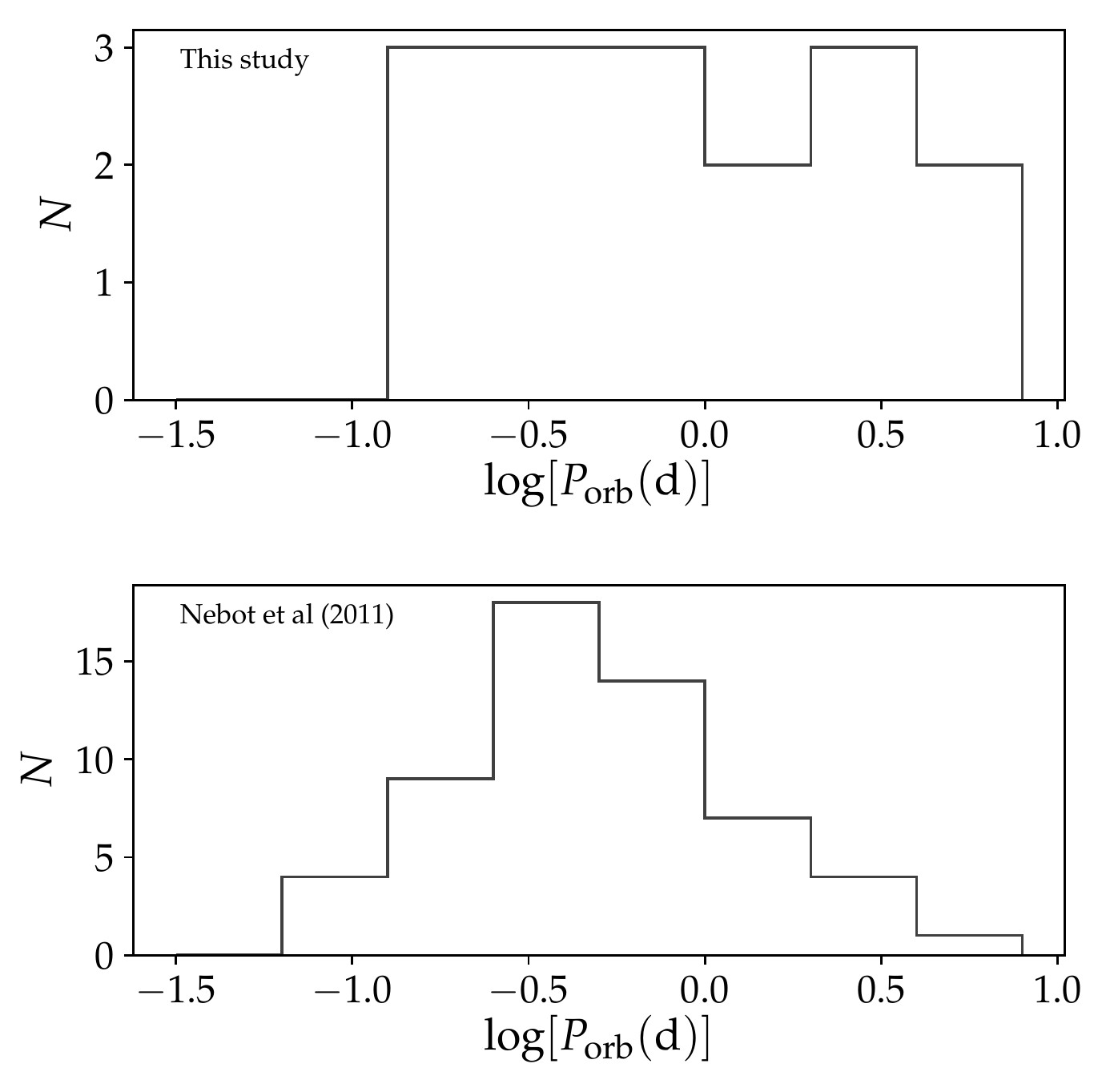}
\caption{The orbital period distribution of the 16 PCEBs contained in this study compared to that of the 79 PCEBs characterised by \citet{Nebot2011}.}
\label{fig:nebot-hist}
\end{figure}

Table \ref{tab:parameters} lists the derived periods for all of the objects. These range from 3.4\,h to 9.2\,d with the centre (the median in $\log P_{\rm orb}$ terms) at $\approx 1$\,d. Since there are only 16 periods in our sample, it was difficult to fit a Gaussian to the period distribution although it has been estimated at around $\log P_{\rm orb}\,(\rm d) = -0.2$ or $P = 0.6$\,d, see Fig. \ref{fig:nebot-hist}.

\begin{figure}
\centering
\includegraphics[width=\columnwidth]{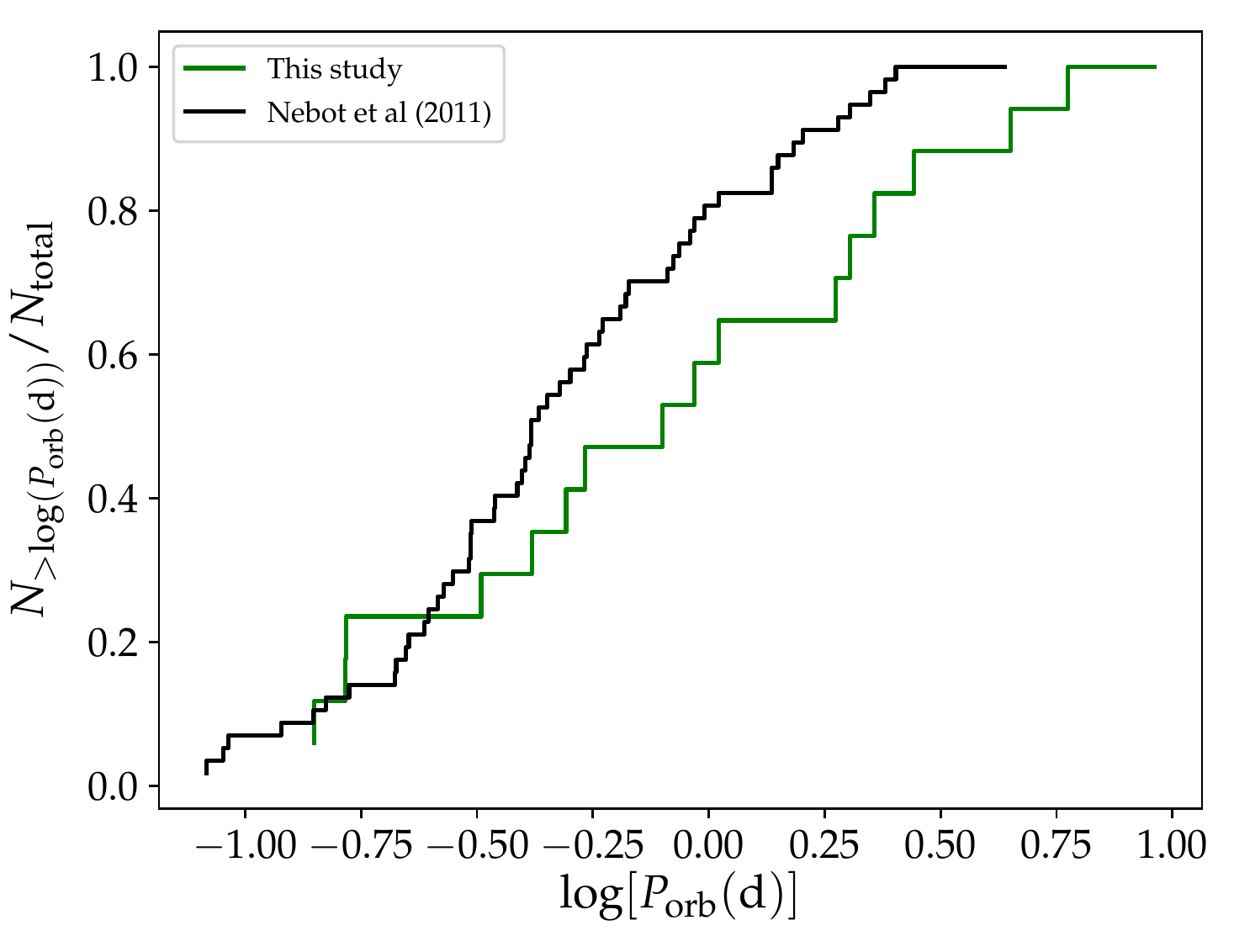}
\caption{Comparison of the cumulative period distribution of this study and the period distribution of 79 PCEBs selected from SDSS by \citet{Nebot2011}. }
\label{fig:nebot-cumulative}
\end{figure}

In a similar study of systems selected from the Sloan Digital Sky Survey (SDSS) undertaken by \citet{Nebot2011}, it was found that the periods range from 1.97\,h to 4.35\,d, and approximately follow a normal distribution in $\log{P_{\rm orb}}$ with a peak centred on $P_{\rm orb}=8.1$\,h or $P_{\rm orb}=0.34$\,d, Fig. \ref{fig:nebot-hist} and \ref{fig:nebot-cumulative} show a comparison between that sample and this one. Their study  had 79 targets, while this only contains 16 with determined periods. We performed an Anderson-Darling test \citep{AndersonDarling} on the two samples to check if they can be considered as being derived from the same underlying population distribution. The results of this test suggest that we cannot reject the null hypothesis i.e. that the two populations are not selected from the same underlying distribution at a 10 per\,cent level.

\begin{figure}
\centering
\includegraphics[width=\columnwidth]{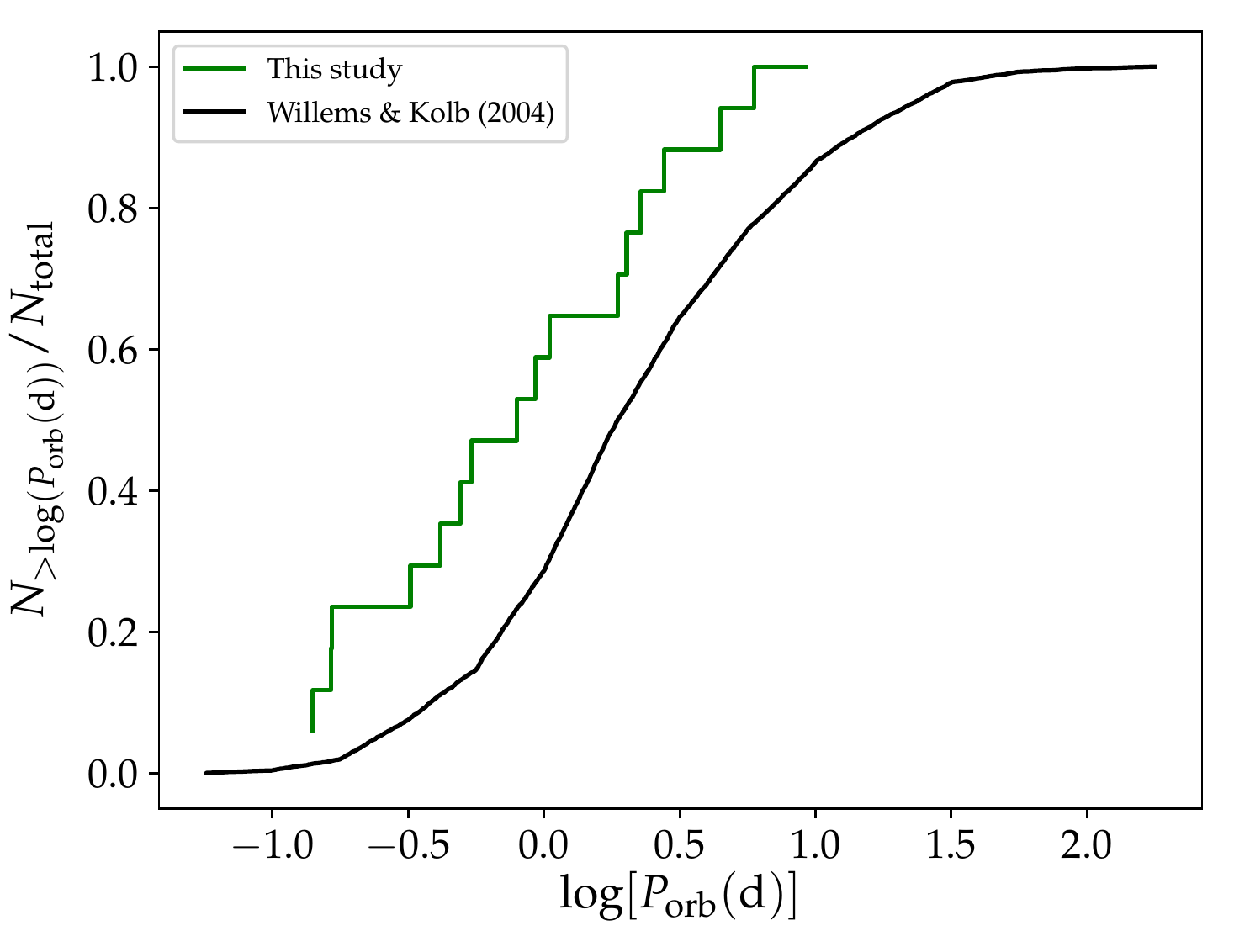}
\caption{Comparison of the cumulative period distribution with a synthetic sample generated from the probability distribution for PCEB periods described in \citet{WillemsKolb}. The Anderson-Darling test suggests the null hypothesis cannot be rejected at a 99 per\,cent confidence level. }
\label{fig:willems-cumulative}
\end{figure}

We also performed an Anderson-Darling test on our sample against the predicted population distribution in \citet{WillemsKolb}. The cumulative distribution function is shown in Fig. \ref{fig:willems-cumulative}. In this case the result suggests we can reject the null hypothesis at a 1 per\,cent level. From this we can conclude that our population does not match the distribution proposed in their BPS and has a peak in period distribution at a shorter duration as can be seen in Fig. \ref{fig:willems-cumulative}. It should be noted that the Willems-Kolb study, adopted a value of $\alpha_\mathrm{CE}=1$. In a more recent study, \citet{Davis2010} produced population distributions for two additional values of $\alpha_\mathrm{CE}$ = 0.1 and 0.6. They find that for low mass secondaries (as per our sample) the theoretical period distribution peaks at $\log P_{\rm orb}\,(\rm d)\approx-0.5$ which is near to our peak at $\log P_{\rm orb}\,(\rm d)\approx-0.2$. \citet{Davis2010} conclude that a value of $\alpha_\mathrm{CE}=0.1$ will predict the observed peak in period distribution, but the distribution will also include a long tail at longer periods going out to hundreds of days. Binaries with such a long period have not yet been observed among PCEB systems and this includes those in the present study. 

More recently \citet{Camacho2014} show that for the PCEBs selected from SDSS, small values for the common envelope efficiency ($\alpha_{\mathrm{CE}}<0.3)$ in simulated populations predict the observed period distribution. Our study, though limited in sample size, agrees with these modelled values of $\alpha_{\mathrm{CE}}$. 

In the \textit{HST} imaging survey of \citet{FarihiFull}, 72 of the 90 targets were confirmed to be WD+MS binaries with the remaining stars deemed to be non-binary. Of the binaries, 43 were resolved with ACS HRC imaging, implying separations wider than at least a few au. The remaining 29 were unresolved with \textit{HST}. We obtained spectroscopy for 20 of the unresolved pairs, of which 16 were confirmed as short-period ($\sim $ hours to days) systems that underwent a common envelope phase. Thus overall, about 60 per\,cent of the binaries have wide orbits and 34 per\,cent are in close orbits with the remaining 6 per\,cent not determined. 

\begin{table}
  \caption{Resulting separation statistics for the sub-samples of the 90 WD+MS binary candidates from \citet{Farihi2006}, classified by separation.}
  \begin{tabular}[width=\textwidth]{ l  c  c }
    \hline
    Type 							 			& Number   & Fraction \\
    \hline
    Candidates 									& 90 		&       \\
    Confirmed as binary 						& 72 	 	& 100\% \\
    Resolved with \textit{HST} (3 -- 1000\,au) 			& 43  		& 60\%  \\
    Unresolved ($\lesssim$\,10 au)			    & 29 		& 40\%  \\
    \hline
    Included in this study 			 			& 20 		& 100\% \\ 
    Confirmed as short period ($\lesssim$10\,au) & 17 		& 85\%  \\
    Number with confirmed periods				& 16 				\\
    Short period suspected, but no solution		& 1                 \\
    Undetected radial velocity variability 		& 3 		& 15\%  \\
    \hline
  \end{tabular}
  \label{tab:populations}
\end{table}

\citet{Nebot2011} calculated a $23\pm2$ per\,cent\, PCEB fraction for WD+MS systems found in SDSS, with an upper limit, after taking selection effects into account, of $27$ per\,cent. In our smaller sample, we obtain a PCEB fraction of $34$ per\,cent, but our sample is biased as it excluded targets that were spatially resolved in ground-based images. Accounting for the fact that spatially-resolved binaries are $63\%$ of the population when viewed from the ground \citep{Farihi2006}, the PCEB fraction of this study reduces to $22$ per\,cent, and is similar to the 23 per\,cent of \citet{Nebot2011}. 

While the difference between the results from this study and that of \citet{Nebot2011} are not statistically significant, it is noted that there is a tendency for slightly longer periods here. \citet{Zorotovic2011} find evidence of a trend towards longer periods for PCEBs with an earlier secondary. Comparison of the secondary spectral types between the Nebot study and this one indicate that there is no discernible difference with both distributions having a median value of approximately M4. Therefore, although this study hints at confirmation of this relationship between secondary spectral type and period, we cannot confirm it conclusively.




\subsection{Observational constraints} \label{constraints}
\begin{figure}
\centering
\includegraphics[width=\columnwidth]{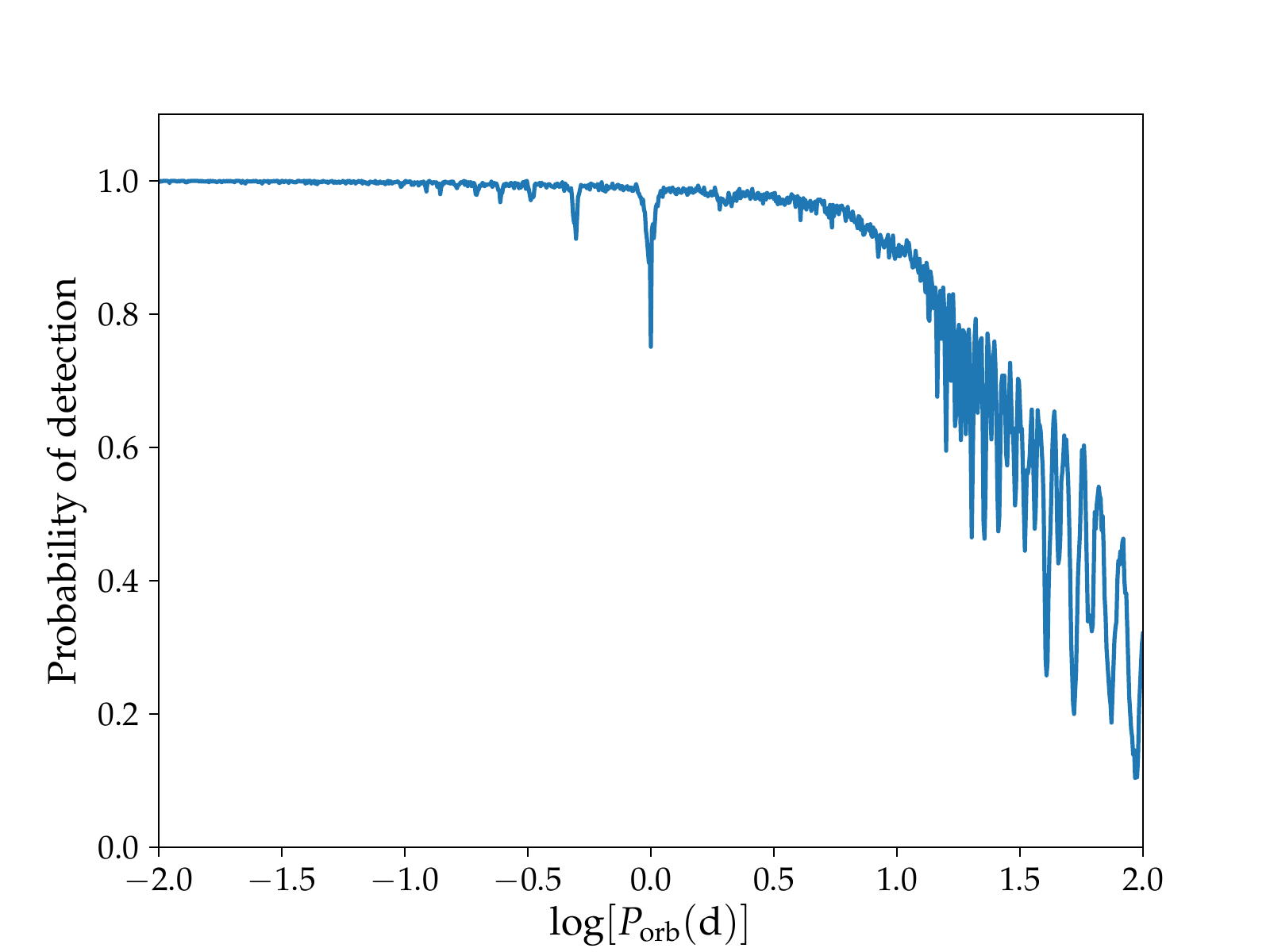}
\caption{The detection probability of radial velocity variations for a simulated input population with periods following a flat distribution in $-2 \leq \log P_{\rm orb} \geq 2$. The detection is modelled on our actual observation times and instrument setup, as described in the text.}
\label{fig:detection}
\end{figure}

In order to understand the effects of our experimental approach in terms of instrument sensitivity and measurement cadence, we performed a Monte-Carlo simulation on a synthetic sample of input period distributions. For each period in our input sample we assumed a random orientation for the inclination of the system with a distribution that is flat in $\cos(i)$. We then calculated a $K_2$ amplitude using the mass function, 
\begin{equation}
	\frac{(M_1 \sin(i))^3}{(M_1+M_2)^2} = \frac{P_\mathrm{orb} K_2^3}{2\pi G} ,
\end{equation}
where $G$ is the gravitational constant. For the masses of the binaries we assumed values of $M_1 = 0.6 \mathrm{M}_{\odot}$ and $M_2 = 0.2 \mathrm{M}_{\odot}$ for all of the objects in the simulation. We randomly selected an HJD based on the actual observing dates and times and computed the simulated observed $K_2$ velocity. We used 15 km\,s$^{-1}$ as the limit on the standard deviation of the simulated velocities to distinguish a detection of radial velocity variation. We repeated this test for 1\,000 input periods to deduce the detection probability as a function of orbital period. The input periods were selected from a random distribution of uniform probability for periods in the range $-2.0 \leq \log P_{\rm orb} (\rm d)\leq 2.0$. The results are shown in  Fig. \ref{fig:detection}. We can be confident that our observing strategy will detect systems with periods ranging from 0.01\,d to $\approx 10$\,d but the detection rate drops of markedly for periods greater than 30\,d. In additional simulations we tweaked two factors; a) the observation baseline, by creating an artificially extended set of observation times and b) an improved observation error, by artificially reducing our measurement error on the wavelength of the spectral features. The former simulation did not lead to improvement in the detect-ability of longer period systems, but the latter significantly improved the chances of finding radial velocities for systems with periods longer than 10\,d. From this we can conclude that our observational setup is limited by the grating resolution and the signal to noise (S/N) of our targets at the telescope and not the time baseline of our observations. 

The three objects that showed no detectable radial velocity variations could have periods that fall into the weeks to several years range and have escaped our detection regime at the INT. Our simulations suggest that improving our radial velocity sensitivity through higher S/N spectra and higher resolution gratings may allow us to detect periods in such systems.

\subsection{Position on the HR diagram}
\begin{figure}
\centering
\includegraphics[width=\columnwidth]{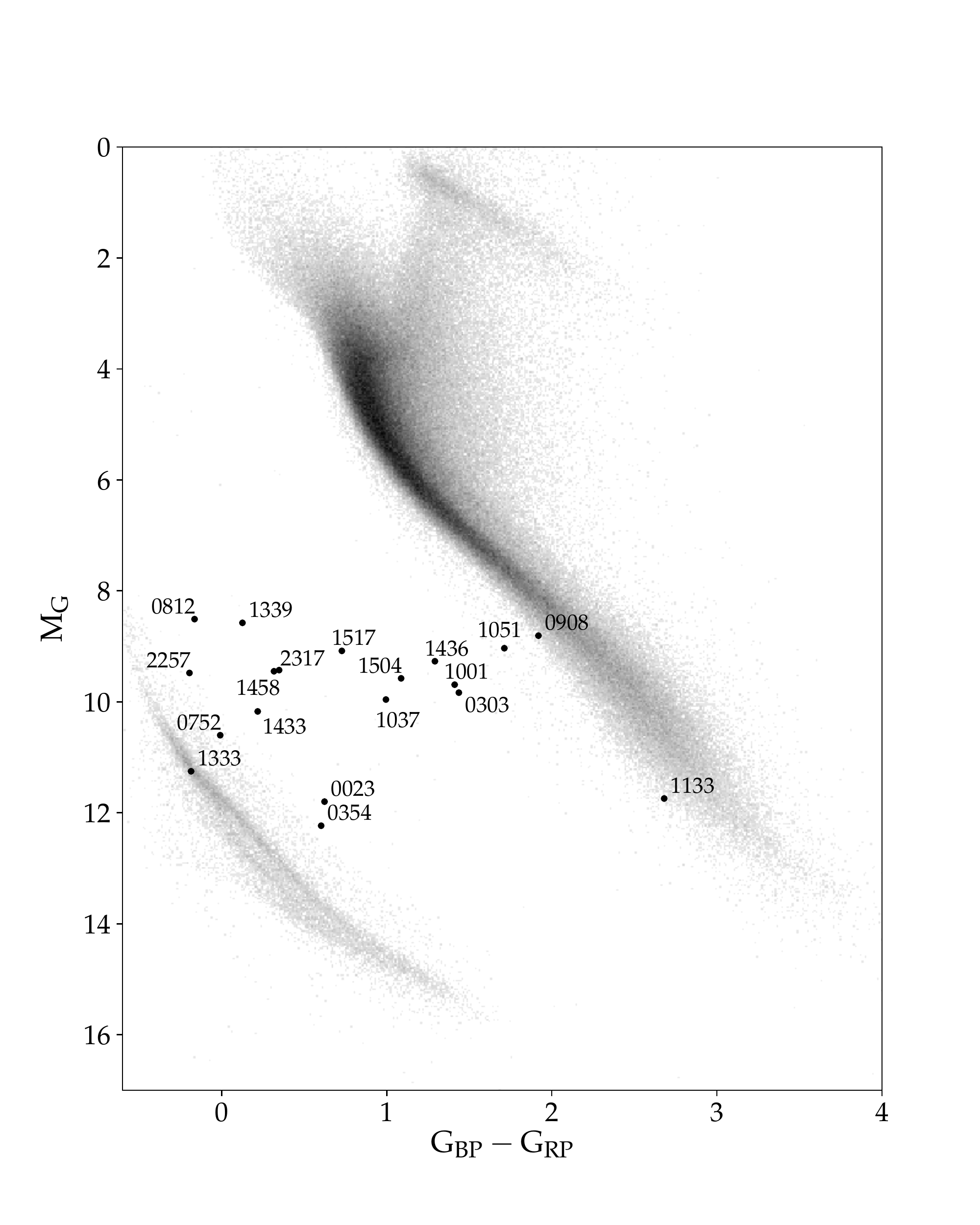}
\caption{Colour-magnitude diagram showing the position of our objects relative to the white dwarf and main sequence. The only member of our sample that is not present in this diagram is WD\,0430+136, which does not have a published parallax in \textit{Gaia} DR2. }
\label{fig:hrdiagram}
\end{figure}

With one exception (WD\,0430+136), all of our targets have \textit{Gaia} measured parallaxes in DR2 \citep{GaiaDR2}. Using these data, we are able to place the binaries on the HR diagram. All points lie between the white dwarf and main sequence as shown in Fig. \ref{fig:hrdiagram}. Some of the objects are at the extreme edges of this range and we discuss those in the following section. 

In Fig. \ref{fig:hrdiagram}, the data used for indicating the overall population of the solar neighbourhood were drawn from a random sample of 327,000 entries in DR2 that matched the quality criterion that the G magnitude was greater than 20 times the error in the \textit{G} magnitude. To increase the visibility of the white dwarf sequence we supplemented the sample with a selected set of 14,000 white dwarfs drawn from DR2 by colour selection.  

\section{Notes on individual objects}\label{individualnotes}

\begin{figure}
\centering
\includegraphics[width=\columnwidth]{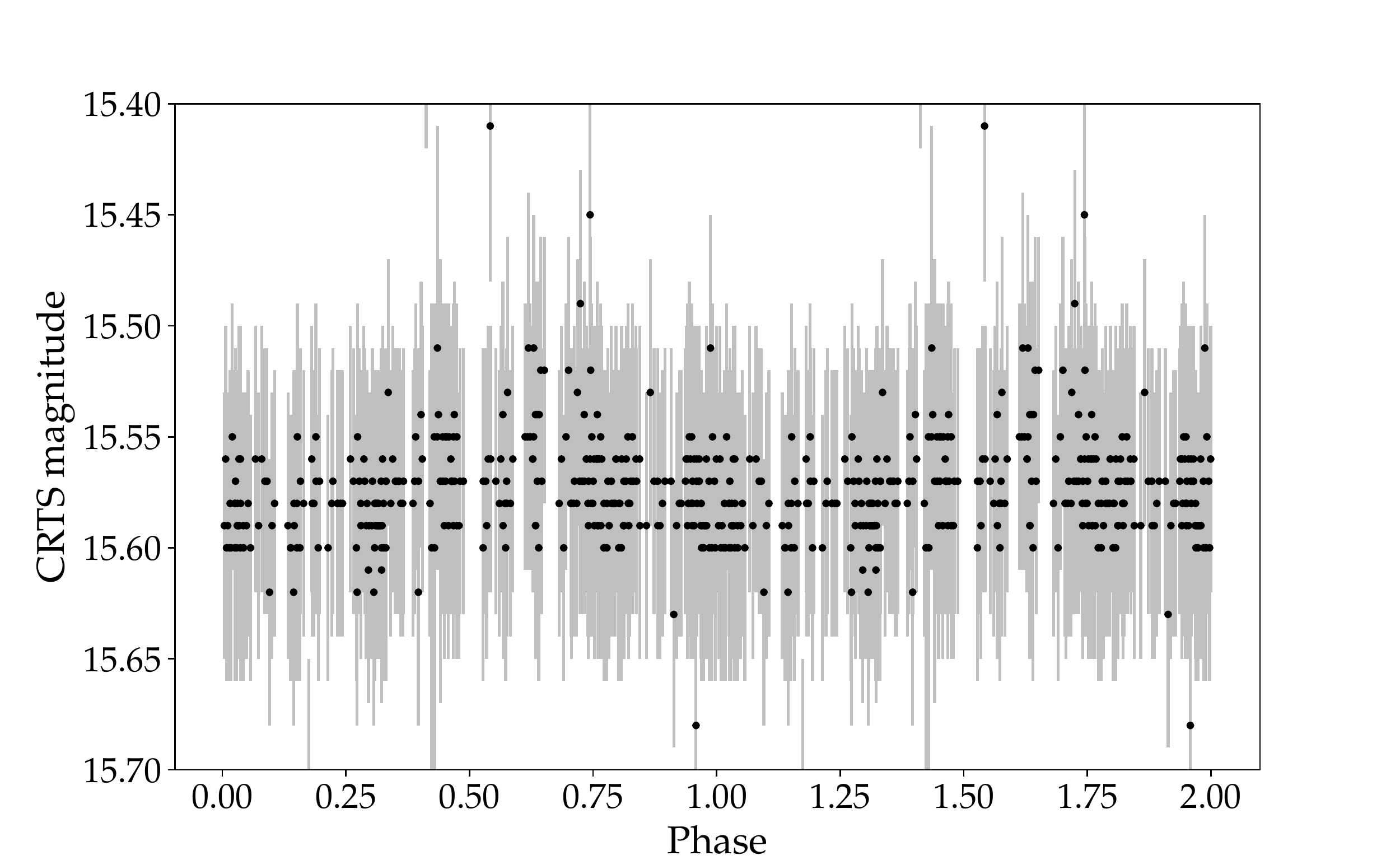}
\caption{The phased-folded CRTS light-curve of WD\,0303-007. At phase 0.5 there is a tentative detection of a reflection effect on the secondary.}
\label{fig:CRTS_0303-007_phased}
\end{figure}

\emph{0303-007}.
Phase folding the CRTS data to our radial velocity derived period, reveals a tentative increase in brightness by about 0.05 magnitude at phase 0.5 as shown in Fig. \ref{fig:CRTS_0303-007_phased}. This could be due to a small reflection effect on the secondary. 

%

\emph{0354+463}.
\citet{2003ApJ...596..477Z} identify this target as containing a DAZ and report that the velocity of the \ion{Ca}{ii} K absorption in the white dwarf and the H\,$\beta$ emission line in the M dwarf agree reasonably well at $-69.7$ km\,s$^{-1}$ and $-67.7$ km\,s$^{-1}$ respectively.  We measured a $\gamma_2$ velocity of $-67$ km\,s$^{-1}$ using the \ion{Na}{i} absorption in the secondary. The metal pollution of the white dwarf atmosphere is likely due to accretion from a stellar wind emanating from the secondary.

\emph{0430+136}.
This is a suspected triple, with one component resolved by \textit{HST} and a second identified on the basis of \textit{I}\,-band photometric excess at the location of the white dwarf \citep{FarihiFull}.  We took 30 spectra of this target over a 1.2 year period and can see no obvious variability above our measurement errors. However, the two components resolved by HST are separated by 0".26 which is several times smaller than typical seeing at Roque de Los Muchachos.  Thus, it is suspected that the Na I line detected in the INT spectra originates in the brighter, widely separated, red dwarf component. If this target is followed up, then it would be more prudent to search for radial velocity variations in the H\,$\alpha$ emission expected from the illuminated surface of the fainter red dwarf that is closer to the white dwarf.

\emph{1133+358}.
This is a rare DC+dM binary \citep{FarihiFull}.  The fact that the pair was unresolved with ACS implies a projected separation of less than 1.3\,au.  Assuming canonical mass values for both of the components an estimate for their separation is 0.06\,au. The white dwarf in this system is likely to be accreting metals from the wind of the M dwarf and this therefore makes it a good target for follow-up investigations looking for metal pollution.

 
\emph{1333+487}.
This was identified as a potential triple system by \citet{FarihiFull} and has a resolved red dwarf companion at a separation of 3\,arcsec with the other component being an unresolved WD+MS binary. We have successfully determined the period of this binary in this study. During our observations, we also took three spectra of the resolved red dwarf. Two spectra were taken within 15 minutes of each other and have radial velocities of 16$\pm1$ and 20$\pm2$\,km\,s$^{-1}$ respectively. The third spectrum was taken 4 days later and has a radial velocity of 0$\pm1$\,km\,s$^{-1}$. There is a possibility that this resolved component is also a binary. Follow-up observations of this object should include spectra of the resolved red dwarf component. The \textit{Gaia} DR2 published proper motions and parallaxes for both components agree to the extent that they are a common proper motion pair, with the small differences explicable in terms of mutual orbital motion. Their parallaxes lead to a minimum separation of $\approx110$\,au.

\emph{1433+538}.
Although \citet{Schultz1996} noted that the H\,$\alpha$ region was featureless during their observations, \citet{FarihiFull} remarked that close inspection of the SDSS spectrum showed some tentative evidence of H\,$\alpha$ and Na 8190\,\AA\ emission that may be due to noise. In this study, we did not find any evidence of emission. 



\begin{figure}
\centering
\includegraphics[width=\columnwidth]{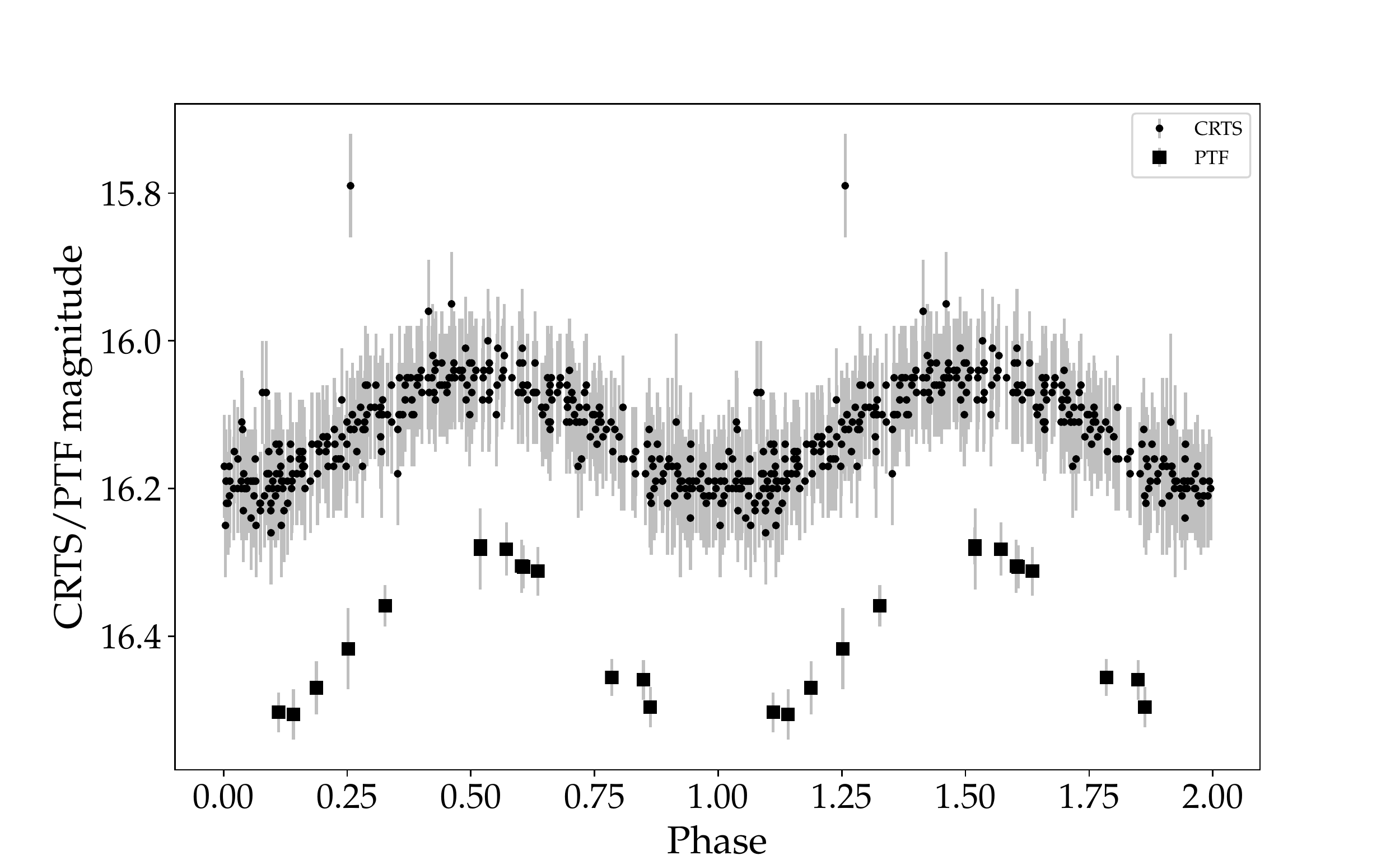}
\caption{The phased-folded light-curve of WD\,1458+171 with photometry taken from CRTS and PTF data. The modulation is attributed to a reflection effect on the secondary. PTF data points are shown with a square symbol and have not been offset from the CRTS data points. It should be noted that CRTS and PTF do not have the same pass band for their photometry. }
\label{fig:1458+171_lc}
\end{figure}

\emph{1458+171}.
\citet{Nebot2011} identify this as a wide binary, based on two spectra taken on the same night approximately 15 min apart. Our Lomb-Scargle analysis has several competing aliases and the frequency for the minimum chi-squared value leads to a period of 0.079\,d. The photometric data from the CRTS and the PTF surveys reveal a clear reflection effect on the secondary and a Lomb-Scargle analysis of these data leads to a best period of 0.164\,d which is double the period of our minimum chi-squared analysis. We have used this longer period as our result. The white dwarf has a temperature of 22 000\,K and with such a short period, a reflection effect is expected. Fig. \ref{fig:1458+171_lc} shows a phase folded plot of the CRTS and PTF data. 

\begin{figure}
\centering
\includegraphics[width=\columnwidth]{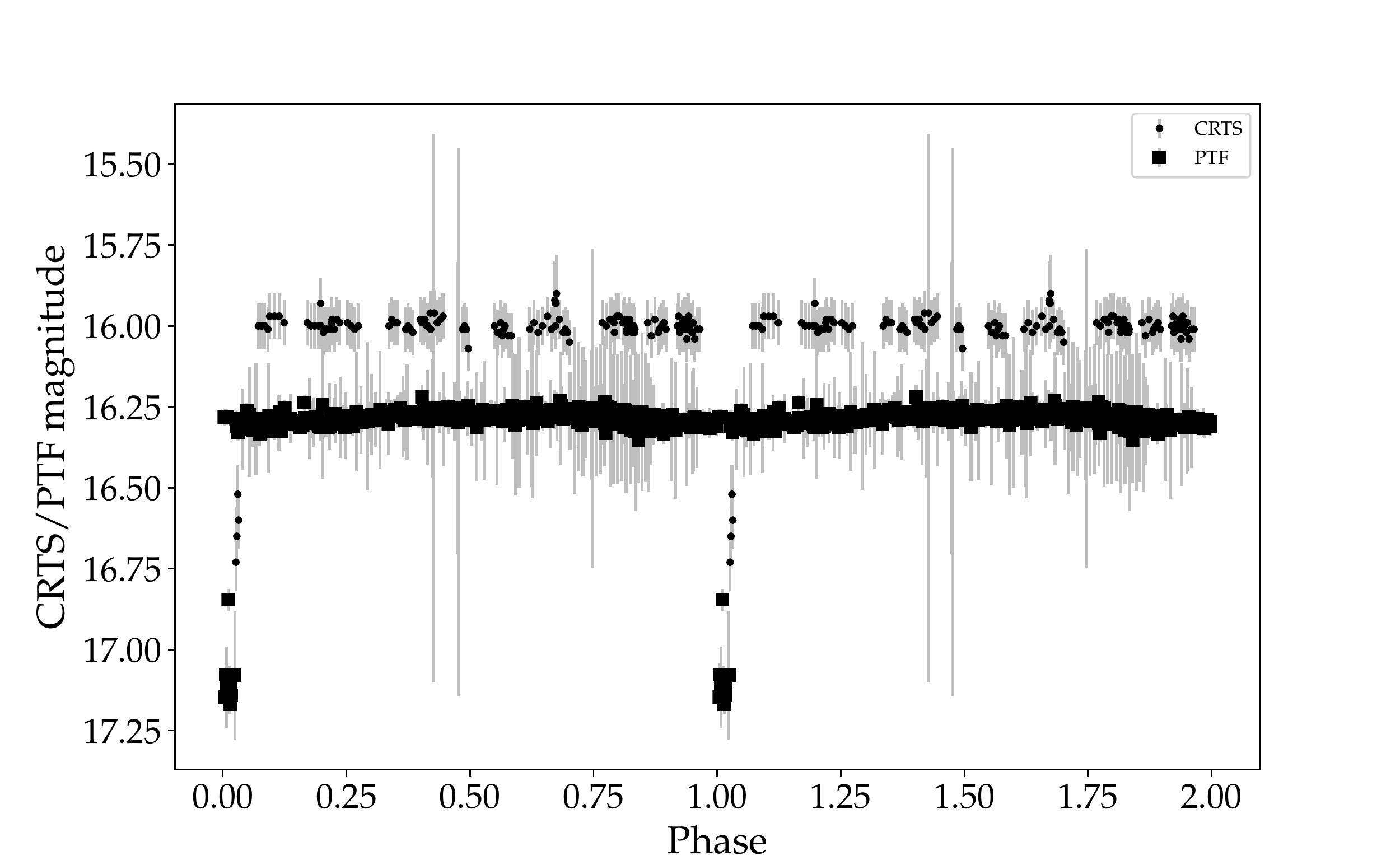}
\caption{The phased-folded light-curve of WD\,1504+546 with photometry taken from the PTF and CRTS surveys. }
\label{fig:1504+546_lc}
\end{figure}

\emph{1504+546}.
The CRTS and PTF photometric data for this target is shown in Fig. \ref{fig:1504+546_lc}. It is an eclipsing binary with a primary eclipse lasting $35$\,min. Although no secondary eclipse is visible there is a reflection effect caused by irradiation of the primary onto the inward facing hemisphere of the secondary. 


\begin{figure}
\centering
\includegraphics[width=\columnwidth]{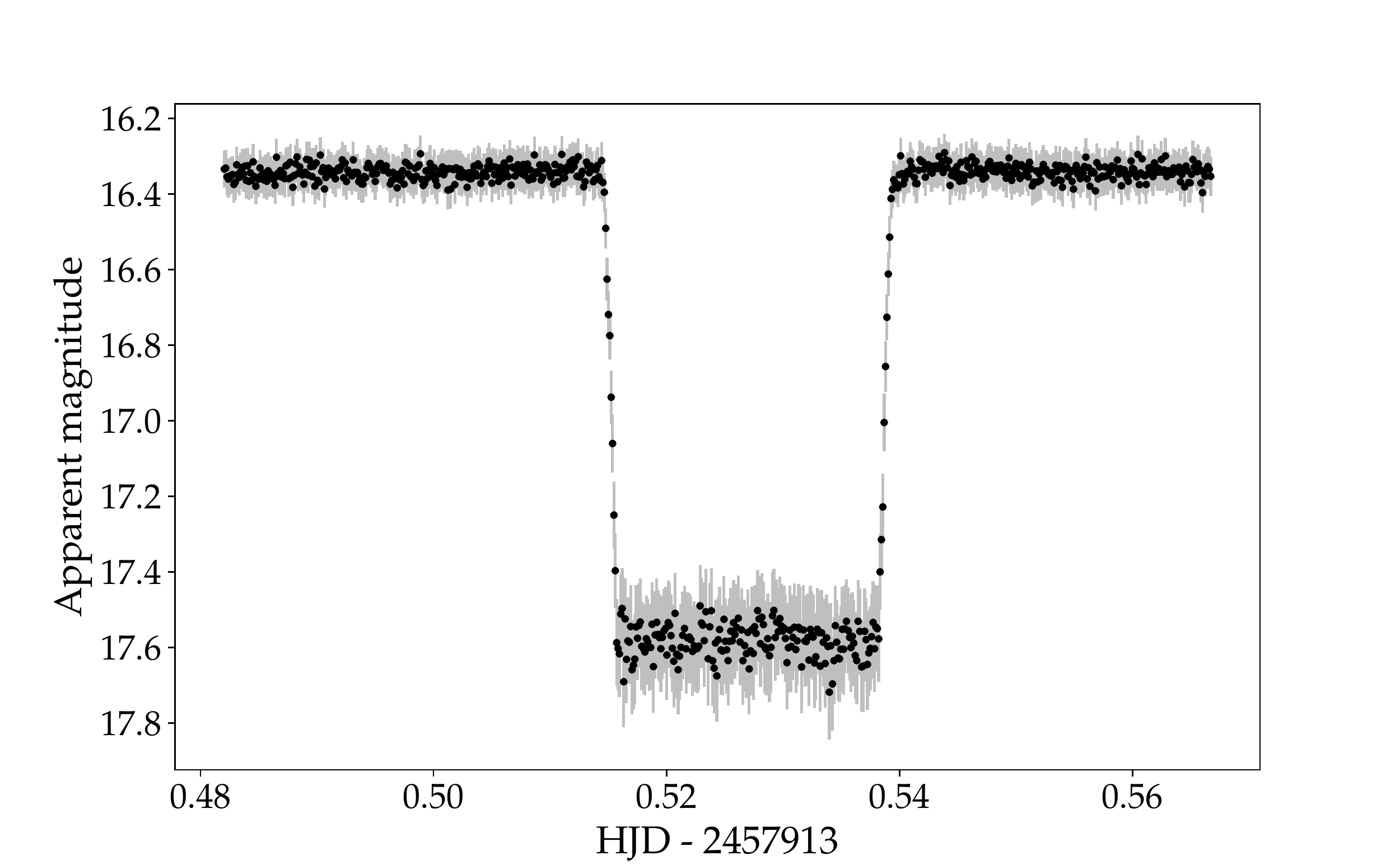}
\caption{The Liverpool RISE camera light curve for the primary eclipse of WD\,1504+546. Integration times were 2\,s and the points plotted here were binned by a factor of 5. }
\label{fig:1504+546_liverpool}
\end{figure}

Photometry of the primary eclipse for WD\,1504+546 was obtained on the Liverpool telescope using the RISE camera with the RISE filter and is shown in Fig. \ref{fig:1504+546_liverpool}. This displays a sharp ingress and egress of the white dwarf behind the secondary with a total eclipse time of $35.205$\,min. Approximate apparent magnitudes were calculated by comparing the flux counts with the published SDSS $r$ band magnitude of a comparison star, SDSS J150618.53+542801.5 located 114 arcseconds to the east of the target. The RISE filter is a broad band $V+R$ filter (covering 5000{\AA} to 8000\,\AA) and so the apparent magnitude is approximate only. 

\emph{1517+502}.
This system was discovered by \citet{Liebert1994} and is a DA white dwarf paired with a dwarf carbon (dC) star. \citet{FarihiFull} failed to resolve the components with the \textit{HST} ACS imaging and set a minimum separation of about 20\,au. Although \citet{Liebert1994} noted a weak Na\,D feature at 5890\,\AA, and this is confirmed in the SDSS spectrum, our spectra did not show any \ion{Na}{i} doublet at 8190\,\AA\, and we were unable to use our fitting method to determine radial velocities. The SDSS spectrum shows clear H\,$\alpha$ emission and this feature might be used in future observations. 


%

\section{Conclusions} 
We have measured the orbital periods and their distribution for 16 WD+MS binaries identified as likely to be PCEB binaries in \citet{FarihiFull}. The periods range from 0.14 to 9.16\,d and the overall distribution is similar to that encountered in previous studies of such populations, \citep{Nebot2011}.  Our sample does not contain any system with a period longer than 9\,d. The longest period in the \citet{Nebot2011} survey is 4.3\,d while the longest known period in a PCEB is that of IK Peg at 21.72\,d \citep{test}. Three of our targets did not show significant radial velocity shifts. It is possible that these targets have periods of around 10 days and longer and therefore the radial velocity amplitudes that are too low to be detected by our particular instrument set up. 

Combining the results of \citet{FarihiFull} and this study, we can confirm the bimodal nature of the period distribution of WD+MS binaries. Recent BPS models \citep{Davis2010} predict a tail of PCEB periods longer than tens of days. Although we see a sharp reduction in the number of systems with these sorts of periods, we are not able to completely rule out the possibility of a small number of systems with periods in this regime. 

When comparing our results to another observational study, \citep{Nebot2011}, we find no statistical difference between our distributions.  Our selection was made by selecting candidates from infrared excess in the 2MASS colours and the SDSS study selected candidates by examining spectra. 


\section*{Acknowledgements}
The research leading to these results has received funding from the European Research Council under the European Union's Seventh Framework Programme (FP/2007-2013) / ERC Grant Agreement n. 320964 (WDTracer). Telescope time was awarded by the PATT allocation of the UK STFC. TRM was supported by the UK STFC grants ST/P000495 and ST/L000733. Data for this paper have been obtained under the International Time Programme of the CCI (International Scientific Committee of the Observatorios de Canarias of the IAC) with the Liverpool Telescope (LT) operated on the island of La Palma in the Observatorio del Roque de los Muchachos.

This work has made use of data from the European Space Agency (ESA) mission
{\it Gaia} (\url{https://www.cosmos.esa.int/gaia}), processed by the {\it Gaia}
Data Processing and Analysis Consortium (DPAC,
\url{https://www.cosmos.esa.int/web/gaia/dpac/consortium}). Funding for the DPAC
has been provided by national institutions, in particular the institutions
participating in the {\it Gaia} Multilateral Agreement.




\bibliographystyle{mnras}
\bibliography{wdms} 

\begin{thebibliography}{}
\makeatletter
\relax
\def\mn@urlcharsother{\let\do\@makeother \do\$\do\&\do\#\do\^\do\_\do\%\do\~}
\def\mn@doi{\begingroup\mn@urlcharsother \@ifnextchar [ {\mn@doi@}
  {\mn@doi@[]}}
\def\mn@doi@[#1]#2{\def\@tempa{#1}\ifx\@tempa\@empty \href
  {http://dx.doi.org/#2} {doi:#2}\else \href {http://dx.doi.org/#2} {#1}\fi
  \endgroup}
\def\mn@eprint#1#2{\mn@eprint@#1:#2::\@nil}
\def\mn@eprint@arXiv#1{\href {http://arxiv.org/abs/#1} {{\tt arXiv:#1}}}
\def\mn@eprint@dblp#1{\href {http://dblp.uni-trier.de/rec/bibtex/#1.xml}
  {dblp:#1}}
\def\mn@eprint@#1:#2:#3:#4\@nil{\def\@tempa {#1}\def\@tempb {#2}\def\@tempc
  {#3}\ifx \@tempc \@empty \let \@tempc \@tempb \let \@tempb \@tempa \fi \ifx
  \@tempb \@empty \def\@tempb {arXiv}\fi \@ifundefined
  {mn@eprint@\@tempb}{\@tempb:\@tempc}{\expandafter \expandafter \csname
  mn@eprint@\@tempb\endcsname \expandafter{\@tempc}}}

\bibitem[\protect\citeauthoryear{{Alam} et~al.,}{{Alam}
  et~al.}{2015}]{Sloan2015}
{Alam} S.,  et~al., 2015, \mn@doi [\apjs] {10.1088/0067-0049/219/1/12}, \href
  {http://adsabs.harvard.edu/abs/2015ApJS..219...12A} {219, 12}

\bibitem[\protect\citeauthoryear{{Belczynski}, {Perna}, {Bulik}, {Kalogera},
  {Ivanova}  \& {Lamb}}{{Belczynski} et~al.}{2006}]{2006ApJ...648.1110B}
{Belczynski} K.,  {Perna} R.,  {Bulik} T.,  {Kalogera} V.,  {Ivanova} N.,
  {Lamb} D.~Q.,  2006, \mn@doi [\apj] {10.1086/505169}, \href
  {http://adsabs.harvard.edu/abs/2006ApJ...648.1110B} {648, 1110}

\bibitem[\protect\citeauthoryear{{Camacho}, {Torres}, {Garc{\'\i}a-Berro},
  {Zorotovic}, {Schreiber}, {Rebassa-Mansergas}, {Nebot G{\'o}mez-Mor{\'a}n}
  \& {G{\"a}nsicke}}{{Camacho} et~al.}{2014}]{Camacho2014}
{Camacho} J.,  {Torres} S.,  {Garc{\'\i}a-Berro} E.,  {Zorotovic} M.,
  {Schreiber} M.~R.,  {Rebassa-Mansergas} A.,  {Nebot G{\'o}mez-Mor{\'a}n} A.,
   {G{\"a}nsicke} B.~T.,  2014, \mn@doi [\aap] {10.1051/0004-6361/201323052},
  \href {https://ui.adsabs.harvard.edu/\#abs/2014A&A...566A..86C} {566, A86}

\bibitem[\protect\citeauthoryear{{Cumming}, {Marcy}  \& {Butler}}{{Cumming}
  et~al.}{1999}]{1999ApJ...526..890C}
{Cumming} A.,  {Marcy} G.~W.,   {Butler} R.~P.,  1999, \mn@doi [\apj]
  {10.1086/308020}, \href {http://adsabs.harvard.edu/abs/1999ApJ...526..890C}
  {526, 890}

\bibitem[\protect\citeauthoryear{{Davis}, {Kolb}  \& {Willems}}{{Davis}
  et~al.}{2010}]{Davis2010}
{Davis} P.~J.,  {Kolb} U.,   {Willems} B.,  2010, \mn@doi [\mnras]
  {10.1111/j.1365-2966.2009.16138.x}, \href
  {http://adsabs.harvard.edu/abs/2010MNRAS.403..179D} {403, 179}

\bibitem[\protect\citeauthoryear{{De Marco}}{{De Marco}}{2009}]{DeMarco2009}
{De Marco} O.,  2009, \mn@doi [\pasp] {10.1086/597765}, \href
  {http://adsabs.harvard.edu/abs/2009PASP..121..316D} {121, 316}

\bibitem[\protect\citeauthoryear{{Drake} et~al.,}{{Drake}
  et~al.}{2014}]{CatalinaCatalog}
{Drake} A.~J.,  et~al., 2014, \mn@doi [\apjs] {10.1088/0067-0049/213/1/9},
  \href {http://adsabs.harvard.edu/abs/2014ApJS..213....9D} {213, 9}

\bibitem[\protect\citeauthoryear{{Dufour}, {Blouin}, {Coutu},
  {Fortin-Archambault}, {Thibeault}, {Bergeron}  \& {Fontaine}}{{Dufour}
  et~al.}{2017}]{MontrealDB}
{Dufour} P.,  {Blouin} S.,  {Coutu} S.,  {Fortin-Archambault} M.,  {Thibeault}
  C.,  {Bergeron} P.,   {Fontaine} G.,  2017, in {Tremblay} P.-E.,  {Gaensicke}
  B.,   {Marsh} T.,  eds,  Astronomical Society of the Pacific Conference
  Series Vol. 509, 20th European White Dwarf Workshop. p.~3 (\mn@eprint {arXiv}
  {1610.00986})

\bibitem[\protect\citeauthoryear{Eggleton \& Tokovinin}{Eggleton \&
  Tokovinin}{2008}]{binaryfraction}
Eggleton P.~P.,  Tokovinin A.~A.,  2008, Monthly Notices of the Royal
  Astronomical Society, 389, 869

\bibitem[\protect\citeauthoryear{{Farihi}, {Hoard}  \& {Wachter}}{{Farihi}
  et~al.}{2006}]{Farihi2006}
{Farihi} J.,  {Hoard} D.~W.,   {Wachter} S.,  2006, \mn@doi [\apj]
  {10.1086/504683}, \href {http://adsabs.harvard.edu/abs/2006ApJ...646..480F}
  {646, 480}

\bibitem[\protect\citeauthoryear{{Farihi}, {Hoard}  \& {Wachter}}{{Farihi}
  et~al.}{2010}]{FarihiFull}
{Farihi} J.,  {Hoard} D.~W.,   {Wachter} S.,  2010, \mn@doi [\apjs]
  {10.1088/0067-0049/190/2/275}, \href
  {http://adsabs.harvard.edu/abs/2010ApJS..190..275F} {190, 275}

\bibitem[\protect\citeauthoryear{{Gaia Collaboration}, {Brown}, {Vallenari},
  {Prusti}, {de Bruijne}, {Babusiaux}  \& {Bailer-Jones}}{{Gaia Collaboration}
  et~al.}{2018}]{GaiaDR2}
{Gaia Collaboration} {Brown} A.~G.~A.,  {Vallenari} A.,  {Prusti} T.,  {de
  Bruijne} J.~H.~J.,  {Babusiaux} C.,   {Bailer-Jones} C.~A.~L.,  2018,
  preprint, \href {http://adsabs.harvard.edu/abs/2018arXiv180409365G} {}
  (\mn@eprint {arXiv} {1804.09365})

\bibitem[\protect\citeauthoryear{{Han} \& {Podsiadlowski}}{{Han} \&
  {Podsiadlowski}}{2004}]{2004MNRAS.350.1301H}
{Han} Z.,  {Podsiadlowski} P.,  2004, \mn@doi [\mnras]
  {10.1111/j.1365-2966.2004.07713.x}, \href
  {http://adsabs.harvard.edu/abs/2004MNRAS.350.1301H} {350, 1301}

\bibitem[\protect\citeauthoryear{{Ivanova} et~al.,}{{Ivanova}
  et~al.}{2013}]{CEreview}
{Ivanova} N.,  et~al., 2013, \mn@doi [\aapr] {10.1007/s00159-013-0059-2}, \href
  {http://adsabs.harvard.edu/abs/2013A%26ARv..21...59I} {21, 59}

\bibitem[\protect\citeauthoryear{{Landsman}}{{Landsman}}{1993}]{test}
{Landsman} W.,  1993, \mn@doi [\pasp] {10.1086/133242}, \href
  {http://adsabs.harvard.edu/abs/1993PASP..105..841L} {105, 841}

\bibitem[\protect\citeauthoryear{{Law} et~al.,}{{Law}
  et~al.}{2009}]{PTFCatalog}
{Law} N.~M.,  et~al., 2009, \mn@doi [\pasp] {10.1086/648598}, \href
  {http://adsabs.harvard.edu/abs/2009PASP..121.1395L} {121, 1395}

\bibitem[\protect\citeauthoryear{{Liebert}, {Schmidt}, {Lesser}, {Stepanian},
  {Lipovetsky}, {Chaffe}, {Foltz}  \& {Bergeron}}{{Liebert}
  et~al.}{1994}]{Liebert1994}
{Liebert} J.,  {Schmidt} G.~D.,  {Lesser} M.,  {Stepanian} J.~A.,  {Lipovetsky}
  V.~A.,  {Chaffe} F.~H.,  {Foltz} C.~B.,   {Bergeron} P.,  1994, \mn@doi
  [\apj] {10.1086/173685}, \href
  {http://adsabs.harvard.edu/abs/1994ApJ...421..733L} {421, 733}

\bibitem[\protect\citeauthoryear{{Liebert}, {Bergeron}  \& {Holberg}}{{Liebert}
  et~al.}{2005}]{Liebert2005}
{Liebert} J.,  {Bergeron} P.,   {Holberg} J.~B.,  2005, \mn@doi [\apjs]
  {10.1086/425738}, \href {http://adsabs.harvard.edu/abs/2005ApJS..156...47L}
  {156, 47}

\bibitem[\protect\citeauthoryear{{Lomb}}{{Lomb}}{1976}]{Lomb1976}
{Lomb} N.~R.,  1976, \mn@doi [\apss] {10.1007/BF00648343}, \href
  {http://adsabs.harvard.edu/abs/1976Ap%26SS..39..447L} {39, 447}

\bibitem[\protect\citeauthoryear{{Marsh}}{{Marsh}}{1989}]{MarshPamela}
{Marsh} T.~R.,  1989, \mn@doi [\pasp] {10.1086/132570}, \href
  {http://adsabs.harvard.edu/abs/1989PASP..101.1032M} {101, 1032}

\bibitem[\protect\citeauthoryear{{Marsh}}{{Marsh}}{1990}]{Marsh1990}
{Marsh} T.~R.,  1990, \mn@doi [\apj] {10.1086/168950}, \href
  {http://adsabs.harvard.edu/abs/1990ApJ...357..621M} {357, 621}

\bibitem[\protect\citeauthoryear{{Marsh}, {Dhillon}  \& {Duck}}{{Marsh}
  et~al.}{1995}]{Marsh1995}
{Marsh} T.~R.,  {Dhillon} V.~S.,   {Duck} S.~R.,  1995, \mn@doi [\mnras]
  {10.1093/mnras/275.3.828}, \href
  {http://adsabs.harvard.edu/abs/1995MNRAS.275..828M} {275, 828}

\bibitem[\protect\citeauthoryear{{Maxted}, {Jeffries}, {Oliveira}, {Naylor}  \&
  {Jackson}}{{Maxted} et~al.}{2008}]{Maxted2008}
{Maxted} P.~F.~L.,  {Jeffries} R.~D.,  {Oliveira} J.~M.,  {Naylor} T.,
  {Jackson} R.~J.,  2008, \mn@doi [\mnras] {10.1111/j.1365-2966.2008.13008.x},
  \href {http://adsabs.harvard.edu/abs/2008MNRAS.385.2210M} {385, 2210}

\bibitem[\protect\citeauthoryear{{McCook} \& {Sion}}{{McCook} \&
  {Sion}}{1999}]{McCookSion1999}
{McCook} G.~P.,  {Sion} E.~M.,  1999, \mn@doi [\apjs] {10.1086/313186}, \href
  {http://adsabs.harvard.edu/abs/1999ApJS..121....1M} {121, 1}

\bibitem[\protect\citeauthoryear{{Morales-Rueda}, {Maxted}, {Marsh}, {North}
  \& {Heber}}{{Morales-Rueda} et~al.}{2003}]{Morales2003}
{Morales-Rueda} L.,  {Maxted} P.~F.~L.,  {Marsh} T.~R.,  {North} R.~C.,
  {Heber} U.,  2003, \mn@doi [\mnras] {10.1046/j.1365-8711.2003.06088.x}, \href
  {http://adsabs.harvard.edu/abs/2003MNRAS.338..752M} {338, 752}

\bibitem[\protect\citeauthoryear{{Nebot G{\'o}mez-Mor{\'a}n} et~al.,}{{Nebot
  G{\'o}mez-Mor{\'a}n} et~al.}{2011}]{Nebot2011}
{Nebot G{\'o}mez-Mor{\'a}n} A.,  et~al., 2011, \mn@doi [\aap]
  {10.1051/0004-6361/201117514}, \href
  {http://adsabs.harvard.edu/abs/2011A%26A...536A..43N} {536, A43}

\bibitem[\protect\citeauthoryear{{Nelemans} \& {Tout}}{{Nelemans} \&
  {Tout}}{2005}]{Nelemans2005}
{Nelemans} G.,  {Tout} C.~A.,  2005, \mn@doi [\mnras]
  {10.1111/j.1365-2966.2004.08496.x}, \href
  {http://adsabs.harvard.edu/abs/2005MNRAS.356..753N} {356, 753}

\bibitem[\protect\citeauthoryear{{Rebassa-Mansergas}
  et~al.,}{{Rebassa-Mansergas} et~al.}{2012}]{2012MNRAS.423..320R}
{Rebassa-Mansergas} A.,  et~al., 2012, \mn@doi [\mnras]
  {10.1111/j.1365-2966.2012.20880.x}, \href
  {http://adsabs.harvard.edu/abs/2012MNRAS.423..320R} {423, 320}

\bibitem[\protect\citeauthoryear{{Scargle}}{{Scargle}}{1982}]{Scargle1982}
{Scargle} J.~D.,  1982, \mn@doi [\apj] {10.1086/160554}, \href
  {http://adsabs.harvard.edu/abs/1982ApJ...263..835S} {263, 835}

\bibitem[\protect\citeauthoryear{Scholz \& Stephens}{Scholz \&
  Stephens}{1987}]{AndersonDarling}
Scholz F.~W.,  Stephens M.~A.,  1987, \mn@doi [Journal of the American
  Statistical Association] {10.1080/01621459.1987.10478517}, 82, 918

\bibitem[\protect\citeauthoryear{{Schreiber}, {G{\"a}nsicke}, {Southworth},
  {Schwope}  \& {Koester}}{{Schreiber} et~al.}{2008}]{Schreiber2008}
{Schreiber} M.~R.,  {G{\"a}nsicke} B.~T.,  {Southworth} J.,  {Schwope} A.~D.,
  {Koester} D.,  2008, \mn@doi [\aap] {10.1051/0004-6361:20078765}, \href
  {http://adsabs.harvard.edu/abs/2008A%26A...484..441S} {484, 441}

\bibitem[\protect\citeauthoryear{{Schreiber} et~al.,}{{Schreiber}
  et~al.}{2010}]{Schreiber2010}
{Schreiber} M.~R.,  et~al., 2010, \mn@doi [\aap] {10.1051/0004-6361/201013990},
  \href {http://adsabs.harvard.edu/abs/2010A%26A...513L...7S} {513, L7}

\bibitem[\protect\citeauthoryear{{Schultz}, {Zuckerman}  \&
  {Becklin}}{{Schultz} et~al.}{1996}]{Schultz1996}
{Schultz} G.,  {Zuckerman} B.,   {Becklin} E.~E.,  1996, \mn@doi [\apj]
  {10.1086/176979}, \href {http://adsabs.harvard.edu/abs/1996ApJ...460..402S}
  {460, 402}

\bibitem[\protect\citeauthoryear{{Tappert}, {G{\"a}nsicke}, {Zorotovic},
  {Toledo}, {Southworth}, {Papadaki}  \& {Mennickent}}{{Tappert}
  et~al.}{2009}]{Tappert2009}
{Tappert} C.,  {G{\"a}nsicke} B.~T.,  {Zorotovic} M.,  {Toledo} I.,
  {Southworth} J.,  {Papadaki} C.,   {Mennickent} R.~E.,  2009, \mn@doi [\aap]
  {10.1051/0004-6361/200912049}, \href
  {http://adsabs.harvard.edu/abs/2009A%26A...504..491T} {504, 491}

\bibitem[\protect\citeauthoryear{{Tremblay} \& {Bergeron}}{{Tremblay} \&
  {Bergeron}}{2007}]{Tremblay2007}
{Tremblay} P.-E.,  {Bergeron} P.,  2007, \mn@doi [\apj] {10.1086/511330}, \href
  {http://adsabs.harvard.edu/abs/2007ApJ...657.1013T} {657, 1013}

\bibitem[\protect\citeauthoryear{{Wachter}, {Hoard}, {Hansen}, {Wilcox},
  {Taylor}  \& {Finkelstein}}{{Wachter} et~al.}{2003}]{Wachter2003}
{Wachter} S.,  {Hoard} D.~W.,  {Hansen} K.~H.,  {Wilcox} R.~E.,  {Taylor}
  H.~M.,   {Finkelstein} S.~L.,  2003, \mn@doi [\apj] {10.1086/367821}, \href
  {http://adsabs.harvard.edu/abs/2003ApJ...586.1356W} {586, 1356}

\bibitem[\protect\citeauthoryear{{Willems} \& {Kolb}}{{Willems} \&
  {Kolb}}{2004}]{WillemsKolb}
{Willems} B.,  {Kolb} U.,  2004, \mn@doi [\aap] {10.1051/0004-6361:20040085},
  \href {http://adsabs.harvard.edu/abs/2004A%26A...419.1057W} {419, 1057}

\bibitem[\protect\citeauthoryear{{Zorotovic}, {Schreiber}, {G{\"a}nsicke}  \&
  {Nebot G{\'o}mez-Mor{\'a}n}}{{Zorotovic} et~al.}{2010}]{Zorotovic2010}
{Zorotovic} M.,  {Schreiber} M.~R.,  {G{\"a}nsicke} B.~T.,   {Nebot
  G{\'o}mez-Mor{\'a}n} A.,  2010, \mn@doi [\aap] {10.1051/0004-6361/200913658},
  \href {http://adsabs.harvard.edu/abs/2010A%26A...520A..86Z} {520, A86}

\bibitem[\protect\citeauthoryear{{Zorotovic} et~al.,}{{Zorotovic}
  et~al.}{2011}]{Zorotovic2011}
{Zorotovic} M.,  et~al., 2011, \mn@doi [\aap] {10.1051/0004-6361/201117803},
  \href {https://ui.adsabs.harvard.edu/\#abs/2011A&A...536L...3Z} {536, L3}

\bibitem[\protect\citeauthoryear{{Zorotovic}, {Schreiber},
  {Garc{\'{\i}}a-Berro}, {Camacho}, {Torres}, {Rebassa-Mansergas}  \&
  {G{\"a}nsicke}}{{Zorotovic} et~al.}{2014}]{Zorotovic2014}
{Zorotovic} M.,  {Schreiber} M.~R.,  {Garc{\'{\i}}a-Berro} E.,  {Camacho} J.,
  {Torres} S.,  {Rebassa-Mansergas} A.,   {G{\"a}nsicke} B.~T.,  2014, \mn@doi
  [\aap] {10.1051/0004-6361/201323039}, \href
  {http://adsabs.harvard.edu/abs/2014A%26A...568A..68Z} {568, A68}

\bibitem[\protect\citeauthoryear{{Zuckerman}, {Koester}, {Reid}  \&
  {H{\"u}nsch}}{{Zuckerman} et~al.}{2003}]{2003ApJ...596..477Z}
{Zuckerman} B.,  {Koester} D.,  {Reid} I.~N.,   {H{\"u}nsch} M.,  2003, \mn@doi
  [\apj] {10.1086/377492}, \href
  {http://adsabs.harvard.edu/abs/2003ApJ...596..477Z} {596, 477}

\makeatother
\end{thebibliography}



\appendix
\section{Additional tables}
In the appendix we include a list of nightly observations and a sample of the radial velocity measurements for each object. The full radial velocity data is available online. 
\begin{table}
  \caption{Spectroscopic configurations used in this study. The instrument used was the Intermediate Dispersion Spectrograph (IDS) mounted on the 2.5m Isaac Newton Telescope at the Roque de los Muchachos observatory on the island of La Palma, Spain.}
  \begin{tabular}{ l  l  l  l }
  \hline
  Date & HJD & Grating & Central wavelength \\
  \hline
    2015/02/10 & 2457064 & R831R & 8300 \AA  \\
    2015/02/11 & 2457065 & R831R & 8300 \AA  \\
    2015/02/12 & 2457066 & R831R & 8300 \AA  \\
    2015/02/13 & 2457067 & R831R & 8300 \AA  \\
    2015/02/14 & 2457068 & R831R & 8300 \AA  \\
    2015/02/15 & 2457069 & R831R & 8300 \AA  \\
    2015/02/16 & 2457070 & R831R & 8300 \AA  \\
    2015/02/17 & 2457071 & R831R & 8300 \AA  \\
    2015/09/03 & 2457269 & R831R & 8200 \AA  \\
    2015/09/04 & 2457270 & R831R & 8200 \& 6562 \AA  \\
    2015/09/05 & 2457271 & R831R & 8200 \& 6562 \AA  \\
    2015/09/06 & 2457272 & R831R & 8200 \& 6562 \AA  \\
    2015/09/07 & 2457273 & R831R & 8200 \& 6562 \AA  \\
    2015/09/08 & 2457274 & R831R & 8200 \& 6562 \AA  \\
    2016/02/12 & 2457431 & R831R & 8022 \& 6565 \AA  \\
    2016/02/13 & 2457432 & R831R & 8022 \& 6565 \AA  \\
    2016/02/14 & 2457433 & R831R & 8022 \& 6565 \AA  \\
    2016/02/29 & 2457448 & R831R & 8022 \& 6565 \AA  \\
    2016/03/01 & 2457449 & R831R & 8022 \& 6565 \AA  \\
    2016/03/02 & 2457450 & R831R & 8022 \& 6565 \AA  \\
    2018/06/18 & 2458298 & R831R & 8190 \AA \\
  \hline
  \end{tabular}
  \label{tab:spectroscopy}
\end{table}

\begin{table}
  \caption{Standards used for flux calibration and telluric removal.}
  \begin{tabular}{ l  l  l  l }
  \hline
  Date & Standards observed \\
  \hline
    2015/02/10 & Feige 34, BD+26 2606       \\ 
    2015/02/11 & Feige 34, BD+26 2606       \\ 
    2015/02/12 & HD 19445, HR1342           \\ 
    2015/02/13 & BD+26 260, HR6110, HR1342  \\ 
    2015/02/14 & HR6110, HD 93521, HZ 15    \\ 
    2015/02/15 & HD 84937, HR3958, HD 93521 \\ 
    2015/02/16 & HR6110, HD 93521           \\ 
    2015/02/17 & HD 19445, HR6110, HD 93521 \\ 
    2015/09/03 & Wolf 1346, G191-B2B        \\ 
    2015/09/04 & Wolf 1346, G191-B2B        \\ 
    2015/09/05 & Wolf 1346, G191-B2B        \\ 
    2015/09/06 & Wolf 1346, G191-B2B        \\ 
    2015/09/07 & Wolf 1346, G191-B2B        \\ 
    2015/09/08 & Wolf 1346, G191-B2B        \\ 
    2016/02/12 & G191-B2B, Grw+70 5824      \\ 
    2016/02/13 & G191-B2B, Grw+70 5824      \\ 
    2016/02/14 & G191-B2B, Grw+70 5824      \\ 
    2016/02/29 & G191-B2B, Grw+70 5824      \\ 
    2016/03/01 & G191-B2B                   \\ 
    2016/03/02 & G191-B2B, Grw+70 5824      \\ 
    2018/06/18 & Feige 98, Grw+70 5824, L1512-34 \\
  \hline
  \end{tabular}
  \label{tab:standards}
\end{table}

\section{Radial velocity figures}
\begin{figure*}
\centering
\includegraphics[width=6in]{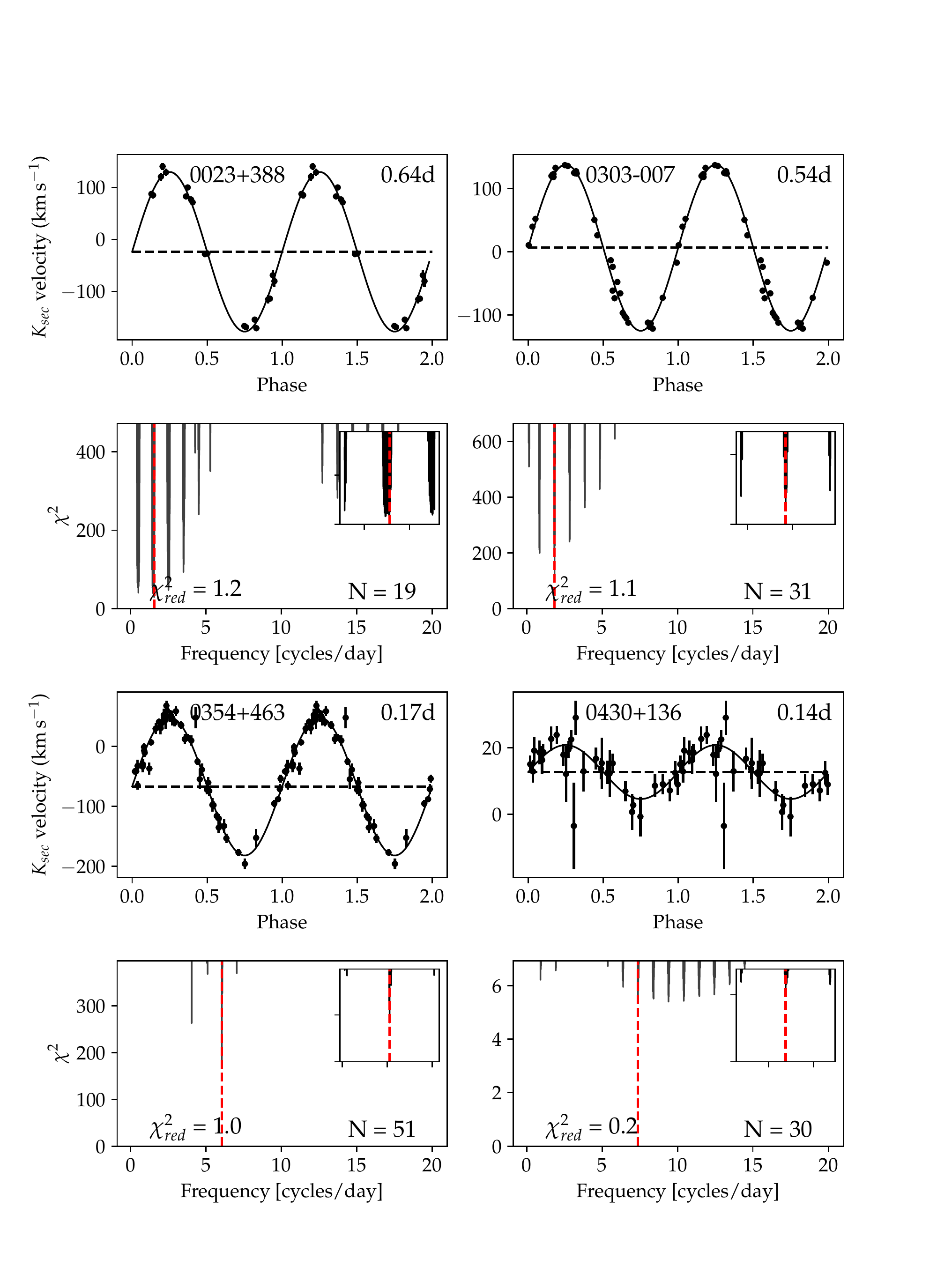}
\caption{Plot of the 19 radial velocity fits. For each object depicted, the upper plot shows the radial velocity in km\,s$^{-1}$ as a function of phase and the lower plot shows the $\chi^2$ (after fitting a sinusoid plus a constant to the data) as a function of frequency . The vertical red line on the periodograms is the frequency chosen as the best fit and the period used for the folded radial velocity curve is derived from this. The insets to the periodograms show a zoomed in region near the best fit period, spanning a region of $\pm 10\%$ around the chosen frequency.}
\label{fig:rvplots}
\end{figure*}

\begin{figure*}
\centering
\includegraphics[width=6in]{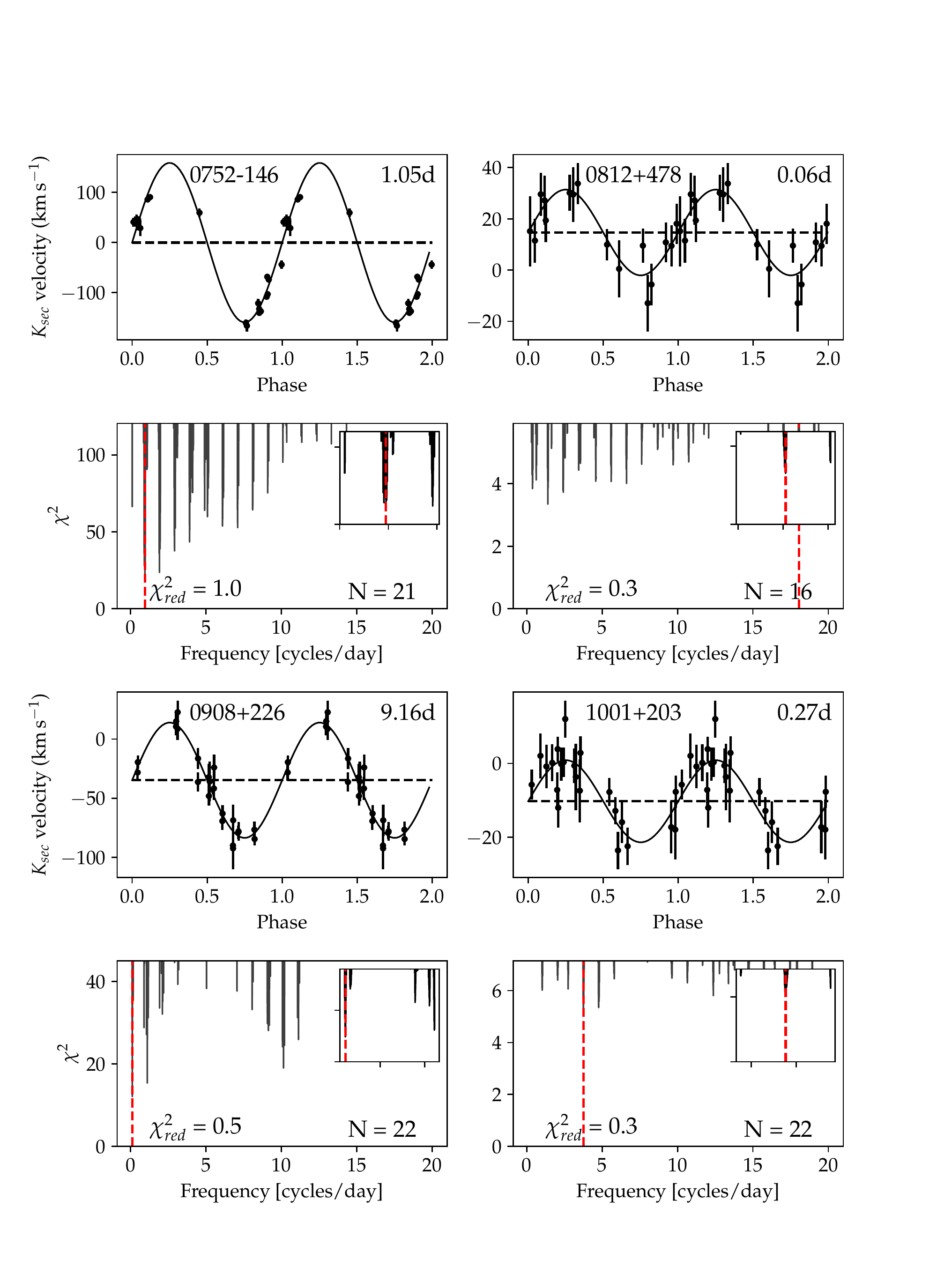}
\contcaption{}
\label{fig:rvplots_2}
\end{figure*}

\begin{figure*}
\centering
\includegraphics[width=6in]{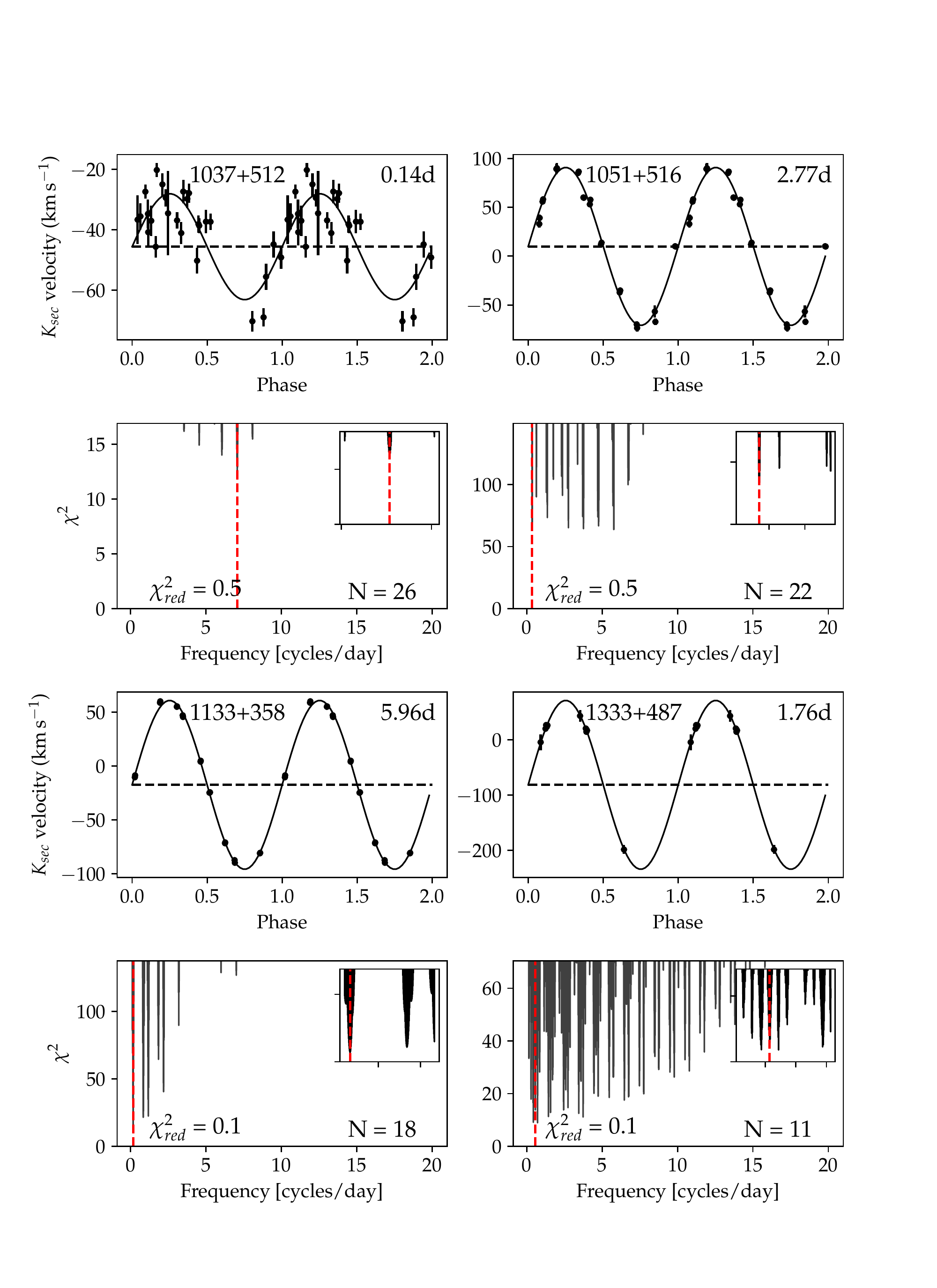}
\contcaption{A blue vertical line is shown where we found a second competing alias has a good chance of being the true period. See notes on individual objects for a discussion.}
\label{fig:rvplots_3}
\end{figure*}

\begin{figure*}
\centering
\includegraphics[width=6in]{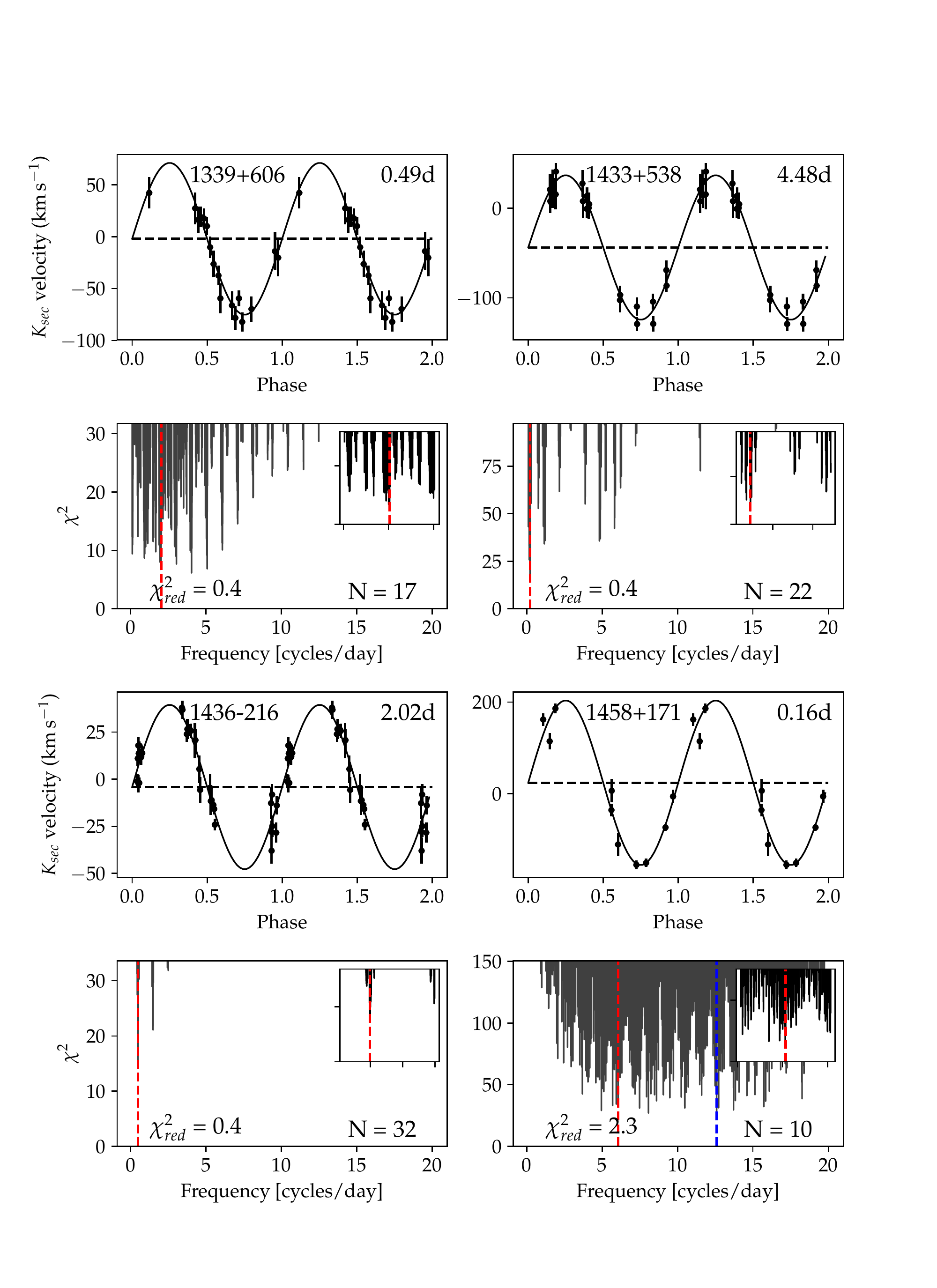}
\contcaption{}
\label{fig:rvplots_4}
\end{figure*}

\begin{figure*}
\centering
\includegraphics[width=6in]{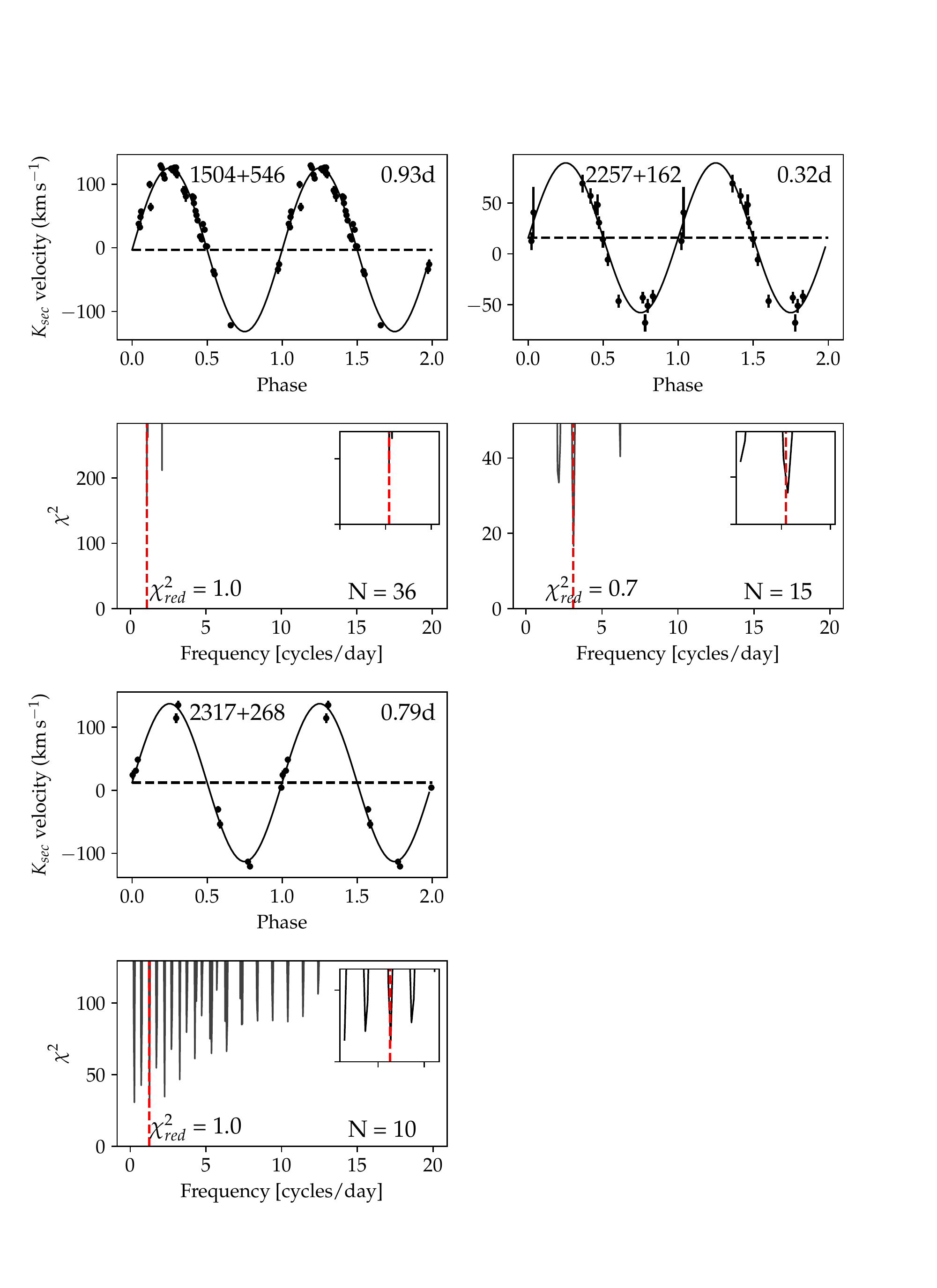}
\contcaption{}
\label{fig:rvplots_5}
\end{figure*}

\pagebreak
\begin{table*}
\caption{The measured radial velocities of our targets from a fit of a double-Gaussian to the \ion{Na}{i} doublet at 8190\AA. This table is abridged. The full data is available at Vizier.}
\begin{tabular}{ l  l  l  l }
	\hline
	WD & HJD & Radial velocity & Error  \\
	   &     & (km\,s$^{-1}$)          & (km\,s$^{-1}$) \\
	\hline
	0023+388 & 2457066.347907 & $-$28.5 & 0.8 \\
	0023+388 & 2457066.358477 & $-$27.2 & 0.9 \\
	0023+388 & 2457269.503428 & 87.3 & 5.1 \\
	0023+388 & 2457269.510507 & 85.0 & 6.2 \\
	0023+388 & 2457270.542192 & $-$166.8 & 3.9 \\
	0023+388 & 2457270.549273 & $-$168.9 & 3.5 \\
	0023+388 & 2457271.597316 & 76.7 & 3.9 \\
	0023+388 & 2457271.604397 & 71.1 & 4.0 \\
	0023+388 & 2457272.511428 & $-$154.5 & 3.9 \\
	0023+388 & 2457272.518510 & $-$171.0 & 4.9 \\
	0023+388 & 2457273.500880 & 82.6 & 4.8 \\
	0023+388 & 2457273.507961 & 99.8 & 4.8 \\
	0023+388 & 2457274.493951 & $-$115.5 & 8.4 \\
	0023+388 & 2457274.501032 & $-$114.2 & 5.7 \\
	0023+388 & 2457432.344926 & $-$69.2 & 10.4 \\
	0023+388 & 2457432.352005 & $-$80.5 & 11.1 \\
	0023+388 & 2458298.655991 & 120.5 & 7.6 \\
	0023+388 & 2458298.663180 & 140.3 & 6.6 \\
	0023+388 & 2458298.677734 & 128.8 & 7.4 \\
	\hline
	0303$-$007 & 2457066.373914 & 137.3 & 3.4 \\
	0303$-$007 & 2457066.384616 & 136.0 & 2.8 \\
	0303$-$007 & 2457067.339107 & 39.7 & 2.0 \\
	0303$-$007 & 2457067.349969 & 52.1 & 2.1 \\
	0303$-$007 & 2457068.350021 & $-$73.0 & 2.2 \\
	0303$-$007 & 2457070.350993 & $-$47.9 & 3.5 \\
	0303$-$007 & 2457070.361546 & $-$65.9 & 3.2 \\
	0303$-$007 & 2457071.350476 & 50.6 & 3.5 \\
	0303$-$007 & 2457071.361029 & 26.1 & 4.7 \\
	0303$-$007 & 2457269.608182 & $-$113.8 & 5.0 \\
	0303$-$007 & 2457269.615262 & $-$121.9 & 4.5 \\
	0303$-$007 & 2457270.602259 & $-$105.6 & 2.6 \\
	0303$-$007 & 2457270.609340 & $-$112.5 & 2.5 \\
	0303$-$007 & 2457271.634996 & $-$61.6 & 2.8 \\
	0303$-$007 & 2457271.642077 & $-$73.7 & 3.0 \\
	0303$-$007 & 2457272.580913 & 124.4 & 2.4 \\
	0303$-$007 & 2457272.587994 & 123.9 & 2.2 \\
	0303$-$007 & 2457272.710581 & $-$13.5 & 2.4 \\
	0303$-$007 & 2457272.717661 & $-$23.9 & 2.1 \\
	0303$-$007 & 2457273.586402 & 118.5 & 2.9 \\
	0303$-$007 & 2457273.593485 & 133.0 & 2.7 \\
  \hline
  \end{tabular}
  \label{tab:rvdata}
\end{table*}



\bsp	
\label{lastpage}
\end{document}